\documentclass[a4paper, oneside, onecolumn, 10pt]{article}
\usepackage[T1]{fontenc}
\usepackage[utf8]{inputenc}
\usepackage{lmodern}

\usepackage{geometry}
\usepackage[skip=0.5\baselineskip, labelfont=bf, singlelinecheck=off]{caption}

\usepackage{enumitem}		      
\usepackage{graphicx}
\usepackage{amsmath}
\usepackage{amssymb}
\usepackage{tensor}             
\usepackage{multicol}         
\usepackage{multirow}
\usepackage{indentfirst}

\usepackage{fancyhdr}
\usepackage{lastpage}
\usepackage{ifthen}
\usepackage{siunitx}
\usepackage{caption}
\usepackage{subcaption}

\usepackage[neverdecrease]{paralist}

\makeatletter
\newcommand{\showfontsize}{\f@size{} pt}
\makeatother

\begin{document}

\setdefaultleftmargin{0.8cm}{0.8cm}{0.8cm}{0.8cm}{0.8cm}{0.8cm}

\baselineskip=5mm

\thispagestyle{fancy}		
\fancyhf{}					
\rhead{RESEARCH ARTICLE}


\lhead{%
\raisebox{-0.5em}
{
\begin{minipage}{100mm}
TRENDS IN SCIENCES, 2024; 21(9): 7653\\
https://doi.org/10.48048/tis.2024.7653
\end{minipage}
}
}

\begin{flushleft}
{\Large\textbf{Characterisation of H$\beta$ spectra from EXO 051910+3737.7 X-rays binaries}}
\end{flushleft}
\vspace{2mm}

\begin{flushleft}
{\large 
\textbf{Pornisara Nuchvanichakul}\textsuperscript{1,2,$\ast$}, 					
\textbf{Puji Irawati}\textsuperscript{3}, 				
\textbf{Pakakaew Rittipruk}\textsuperscript{3}, 	    
\textbf{Poshak Gandhi}\textsuperscript{2},				
\textbf{Christian Knigge}\textsuperscript{2},			
\textbf{Phil Charles}\textsuperscript{2},			    
\textbf{Suwicha Wannawichian}\textsuperscript{4}        
}
\end{flushleft}
\vspace{2mm}

\begin{flushleft}
$^1$\;\textit{M.S. Program in Applied Physics, Department of Physics and Materials Science, Faculty of Science, Chiang Mai University, Chiang Mai 50200, Thailand}\\
$^2$\;\textit{Department of Physics and Astronomy, University of Southampton, Southampton, United Kingdom}\\
$^3$\;\textit{National Astronomical Research Institute of Thailand, Chiang Mai 50180, Thailand}\\
$^4$\;\textit{Department of Physics and Materials Science, Faculty of Science, Chiang Mai University, Chiang Mai 50200, Thailand}
\end{flushleft}
\vspace{2mm}

\begin{flushleft}
\textbf{
($^\ast$\;Corresponding author's e-mail: 
pornisaranuchvanichakul@gmail.com)				
}
\end{flushleft}
\vspace{2mm}

\begin{flushleft}
{\small 
\textit{Received: 26 October 2023}, 
\textit{Revised: 23 November 2023}, 
\textit{Accepted: 20 January 2024}, 
\textit{Published: 10 August 2024}
}
\end{flushleft}
\vspace{3mm}


\begin{flushleft}
\textbf{Abstract}
\end{flushleft}

We investigate the spectral characteristics of the Be/X-ray binary system, EXO 051910+3737.7, in which Be/X-ray systems are the largest sub-class of high-mass X-ray binaries. Spectroscopic observations are taken by the Thai National Telescope (TNT) with a Medium-RESolution spectrograph (MRES) instrument for seven nights spanning from 2020 to 2021. Our primary focus is directed towards the analysis of two Balmer lines, namely H$\alpha$ and H$\beta$, given that Be stars typically exhibit emission features in at least one of these hydrogen Balmer lines during certain phases. Our observations reveal split Balmer emission lines throughout the entire duration of our monitoring. Double Gaussian profiles were employed for line fitting to characterize these lines. The presence of double peaks in the Balmer lines indicates the presence of asymmetries within the circumstellar disc. We then analyze V/R variations and the changes in H$\beta$ spectra. Our analysis of V/R variation which Violet (V) and Red (R) peak intensity components, revealed rapid fluctuations occurring within a single day, although determining the precise periodicity was constrained by instrumental limitations and the duration of observability. Furthermore, employing observed the wavelength differences ($\Delta\lambda$) in conjunction with typical Be star parameters allowed us to estimate the radius ($r_{\beta}$) of the H$\beta$ emitting envelope. The average value was calculated to be 2.585$r_*$, with a standard deviation of 0.050$r_*$.


\section*{\normalsize Introduction}

High-mass X-ray binaries (HMXBs) are binary systems where a massive OB-star is held in orbit by a compact object, such as a neutron star or a black hole. These OB-star companion’s mass is usually more than ten times solar-mass, giving HMXBs remarkable brightness and making them valuable tools for studying the formation of stars and compact celestial objects within our Milky Way galaxy. Estimating the total number of HMXB systems in our galaxy is challenging due to uncertainties related to how these binaries evolve over time. Currently, over a hundred well-documented HMXBs are known to us \cite{Liu2006,Bird2006,Neumann2023}. HMXBs can be divided into three subcategories based on the characteristics of the mass-donating star. There are OB-supergiant systems (SgXRBs), Be/X-ray binaries (BeXRBs), and Roche-lobe overflows (RLOs). Our focus in this work will be on BeXRBs.

\begin{figure}[!h]
\centering
\includegraphics[width=0.7\textwidth]{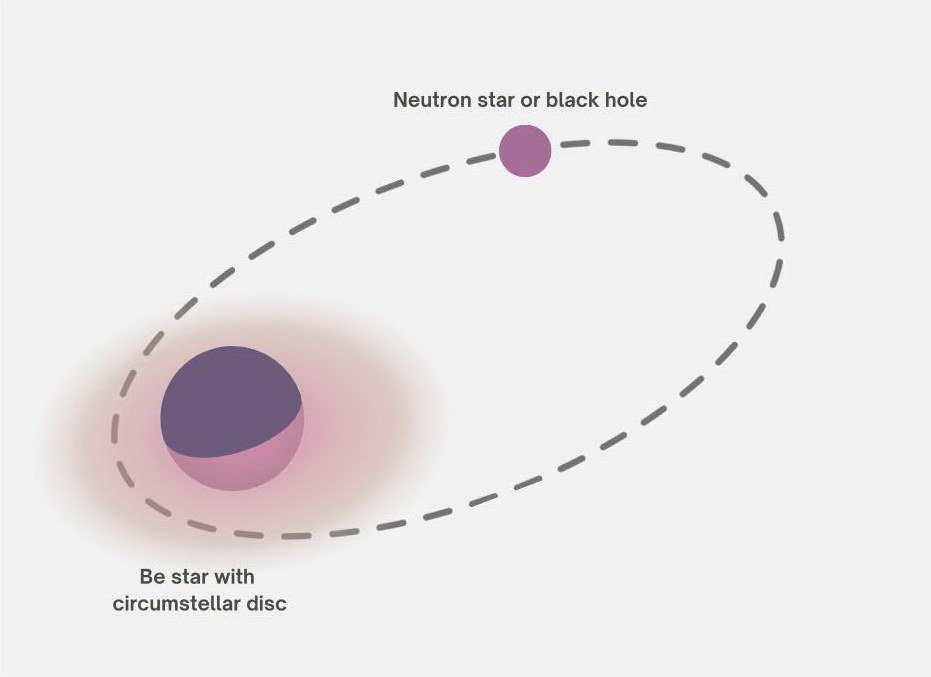}
\caption{Schematics illustrating the model of Be/X-ray binaries, comprising a Be star surrounded by a circumstellar disc and a compact object.}
\label{Fig:be_star}
\end{figure}

The majority of HMXBs fall within BeXRBs. The number of BeXRBs in the Milky Way galaxy is more than 70 systems \cite{Walter2015}. BeXRBs typically have wide and elliptical orbits due to the kick of the supernova. This system consists of a Be star and a compact object, typically a neutron star. The mass range of a massive B star is between 8 and 15 $M_{\odot}$. The mass of the Be star is lost through a circumstellar disc and its stellar wind. These two mass loss mechanisms generate X-ray bursts, commonly occurring when the neutron star approaches the circumstellar disc \cite{OkazakiandNegueruela2001}. The model of BeXRBs is shown in \textbf{Figure~\ref{Fig:be_star}}.

\setcounter{page}{2}
\pagestyle{fancy}		
\fancyhf{}					
\rhead{\thepage\;of\;\pageref{LastPage}}
\lhead{\textit{Trends Sci.} 2024; 21(9): 7653}

In this study, we only observed Be star companions as optical counterparts since compact objects are not observable in optical counterparts. Therefore, our focus is on the characteristics of Be star spectra. Be stars are characterized as B-type stars with at least one of the hydrogen’s Balmer lines emission. These stars rotate rapidly and typically possess a surrounding shell or gas disc. The circumstellar material may originate from mass loss or the accretion of material from an evolved companion star. Be stars are known for their brightness and spectral characteristic variabilities. Therefore, BeXRBs are particularly interesting.
The promising source, EXO 051910+3737.7 (also known as 4U 0515+38 or optical HD 34921), is one of BeXRBs. EXO 051910+3737.7 was classified as B0III—IVpe, with an estimated distance of approximately 1.7$\pm$0.1 kpc \cite{Polcaro1989}. In the recent astrometry observation, $\rm Gaia$ mission, distance of EXO 051910+3737.7 after parallax zero correction is 1.32$\pm$0.05 kpc and the \texttt{G}-magnitude is 7.232$\pm$0.003 \cite{Gaia2022}, making it a great candidate for study due to its brightness.

In a previous study, observed spectra from this binary system revealed variability in the He II $\lambda$4686 emission line. This variability, recognized as a signature of an accretion disc, plays an essential role in identifying HD 34921 as the counterpart to 4U 0515+38 \cite{Polcaro1989}. Concurrently, the emission lines of He II, H$\alpha$, and H$\beta$, were found to vary on short timescales, with a periodicity of order of 300 seconds for the variations of Violet (V) and Red (R) peak intensity \cite{Rossi1991}. Both H$\alpha$ and H$\beta$ exhibited split profiles, consistent with a disc-like structure. Given the analogous variability timescale observed for He II, indicating the presence of an accretion disc, it is inferred that the split profiles of H$\alpha$ and H$\beta$ likewise relate to the vicinity of a compact object \cite{Rossi1991}. 

For 30 years, there have been no optical spectra observed for EXO 051910+3737.7. Therefore, our goal is to present the current spectral analysis of EXO 051910+3737.7 using observations from ground-based telescopes. Additionally, we investigate any possible variations within the Balmer lines, specifically in H$\beta$, which could be used to determine physical parameters such as envelope radius.


\section*{\normalsize Observations and data reductions}

We obtained 22 spectroscopic data of EXO 051910+3737.7 using the 2.4-meter Thai National Telescope (2.4-m TNT) at the Thai National Observatory (TNO) on seven nights between 2020-2021. The observatory is located at 18$^{\circ}$ 34$^{\prime}$ N, 98$^{\circ}$ 28$^{\prime}$ E and 2457 meters above sea level. The spectra were recorded with an Andor 2048 x 512 pixel CCD using a medium-resolution spectrometer (MRES). The MRES has a spectral resolution of 15,000 and spans wavelengths ranging from 390 to 880 nm. The exposure time depends on the magnitude of the objects and the sky condition each night. In order to attain an optimal signal-to-noise ratio (S/N), the exposure times were set within the range of 60 to 240 seconds. Specifically, the higher magnitude of the stars and the cloudy sky relate to lower S/N value and accordingly longer exposure time. The observation details are shown in \textbf{Table~\ref{Table:observation}}.

\begin{table}
\centering
\caption{The observation log to obtain the spectra of EXO 051910+3737.7. The records contain the date, frame number, Heliocentric Julian Date (HJD), exposure time of each frame (in seconds), signal-noise ratio in red and blue parts, and sky condition(s) for each observation.}
\label{Table:observation}
\begin{tabular}{cccccc}
\hline
\textbf{Date} & \textbf{Frame} & \textbf{HJD} & \textbf{Exp. time }            & \textbf{S/N}       & \textbf{Sky condition(s)} \\ 
             &                  &              &  \textbf{(seconds)}         & \textbf{(red/blue)} &  \\
\hline
18--Nov--2020   & 001 & 2459172.31073 & 60      & 78/56   & clear sky   \\
                & 002 & 2459172.31291 & 120     & 143/118 & clear sky   \\
                & 003 & 2459172.37204 & 120     & 161/141 & clear sky  \\
                & 004 & 2459172.44048 & 120     & 131/90  & clear sky  \\
                & 005 & 2459172.44251 & 120     & 127/106 & clear sky  \\
\hline
02--Jan--2021   & 001 & 2459217.13162 & 120     & 42/52   & clear sky  \\
                & 002 & 2459217.13351 & 120     & 56/42   & clear sky   \\
                & 003 & 2459217.13541 & 120     & 82/54   & clear sky  \\
\hline
05--Jan--2021   & 001 & 2459220.01260 & 120     & 81/33   & clear sky  \\
                & 002 & 2459220.01428 & 120     & 59/25   & clear sky  \\
                & 003 & 2459220.10218 & 180     & 25/21   & clear sky  \\
                & 004 & 2459220.10482 & 120     & 7/1    & limited time  \\
                & 005 & 2459220.10658 & 60      & 18/7     &limited time  \\
\hline
16--Mar--2021   & 001 & 2459290.14150 & 120     & 17/11   & cloudy \\
                & 002 & 2459290.14412 & 180     & 48/18   & cloudy  \\
                & 003 & 2459290.14734 & 240     & 74/30   & cloudy  \\
                & 004 & 2459290.17782 & 240     & 27/8    & cloudy  \\
\hline
18--Mar--2021   & 001 & 2459292.04618 & 240     & 158/99  & cloudy, strong wind  \\
\hline
08--Apr--2021   & 001 & 2459313.02591 & 240     & 120/67  & partly clouds  \\
                & 002 & 2459313.05456 & 240     & 81/54   & partly clouds \\
\hline
07--May--2021   & 001 & 2459342.03902 & 240     & 23/14   & cloudy, thunder  \\
                & 002 & 2459342.04412 & 240     & 24/44   & cloudy, thunder	   \\
\hline

\end{tabular}
\end{table}

The spectral data underwent reduction steps, including bias subtraction, flat field correction, and wavelength calibration, performed using standard procedures available in \texttt{IRAF}. ThAr comparison lamp spectra, taken at the beginning and/or at the end of the observing run, were used for the wavelength calibration. Subsequently, continuum normalization was carried out with \texttt{IRAF} by applying the \texttt{CONTINUUM} task for Legendre polynomial fitting to line-free regions in the spectra.

\section*{\normalsize Spectral characteristic}

\begin{figure}
     \centering
     \begin{subfigure}[b]{0.48\textwidth}
         \centering
         \includegraphics[width=\textwidth]{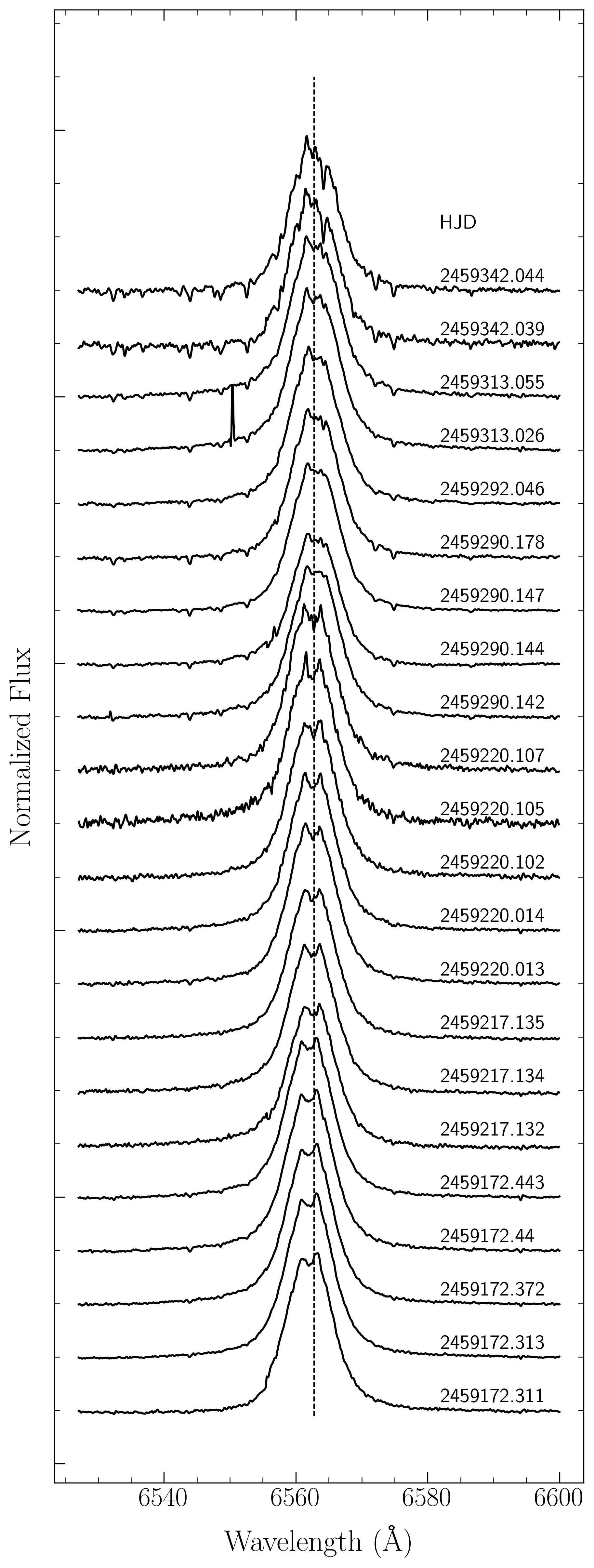}
         \caption{The H$\alpha$ region in the spectrum which dashed line is at 6562.77 \AA.}
         \label{fig:H-alphaorder}
     \end{subfigure}
     \hfill
     \begin{subfigure}[b]{0.48\textwidth}
         \centering
         \includegraphics[width=\textwidth]{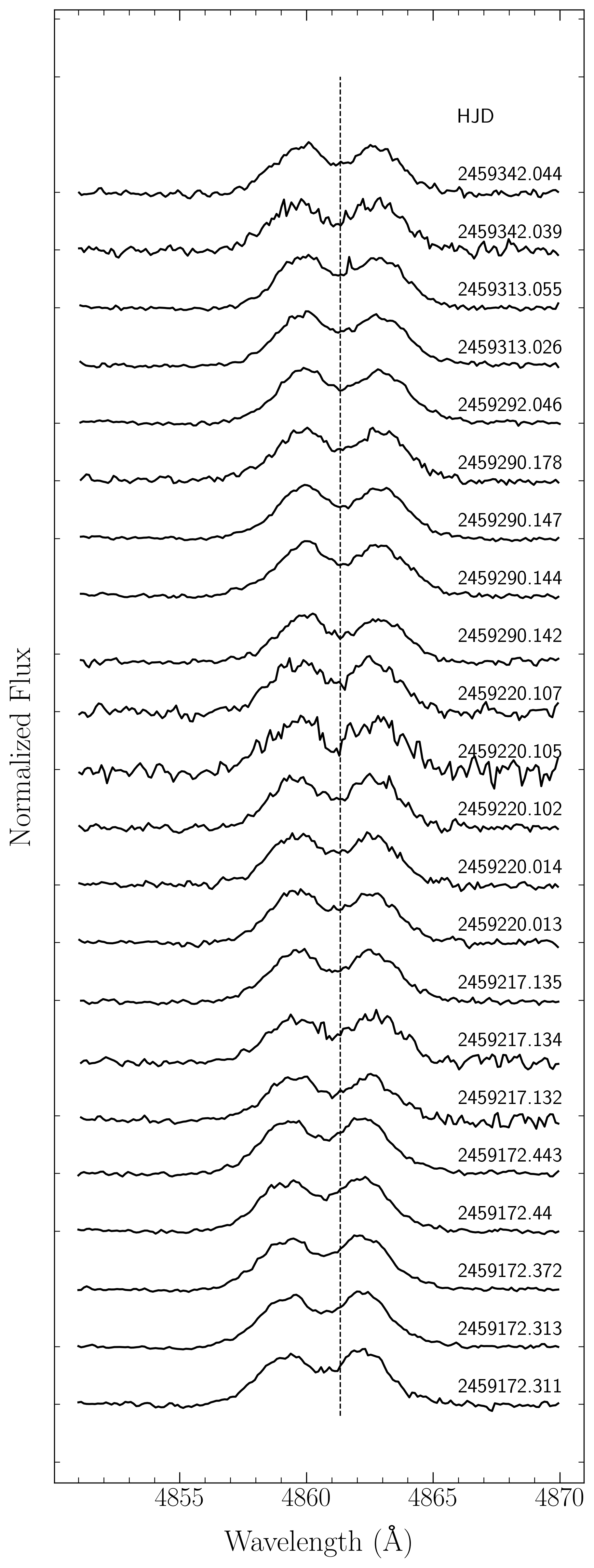}
         \caption{The H$\beta$ region in the spectrum which dashed line is at 4861.33 \AA.}
         \label{fig:H-betaorder}
     \end{subfigure}
        \caption{Samples of Balmer lines were taken by TNO/MRES on entire observations. The Y-axis is normalized flux. The spectra are shifted vertically for easier viewing of the data.}
        \label{Fig:Hbeta_variation}
\end{figure}

Throughout the observations of EXO 051910+3737.7 for seven nights between 2020-2021, we obtained a total of 22 spectra, covering three Balmer lines, H$\alpha$, H$\beta$, and H$\gamma$. Across all spectral observations, there are two promising emission lines, i.e., H$\alpha$, and H$\beta$ as shown in \textbf{Figure~\ref{Fig:Hbeta_variation}}. In \textbf{Figure~\ref{fig:H-alphaorder}}, robust H$\alpha$ emissions consistently reach peak intensities of approximately $\sim$4 times the continuum flux. During the initial observation phase, H$\alpha$ exhibits the separation into two peaks. However, these spectral features are less clear towards the end of the observation, although we took even longer exposure time. 

In \textbf{Figure~\ref{fig:H-betaorder}}, H$\beta$ line has lower peak intensities than H$\alpha$ line. However, there are very interesting features in H$\beta$ lines. All observed spectral lines in H$\beta$ are clearly separated into two peaks. This characteristic allows us to investigate the underlying physical attributes of symmetries in the circumstellar disc and radius of H$\beta$ emitting envelope of Be stars in HMXBs. Nevertheless, the H$\gamma$ regions exhibit a notably low signal-to-noise ratio (S/N). Detailed analysis in this specific area is unfeasible.

\begin{figure}
     \centering
     \begin{subfigure}[b]{0.48\textwidth}
         \centering
         \includegraphics[width=\textwidth]{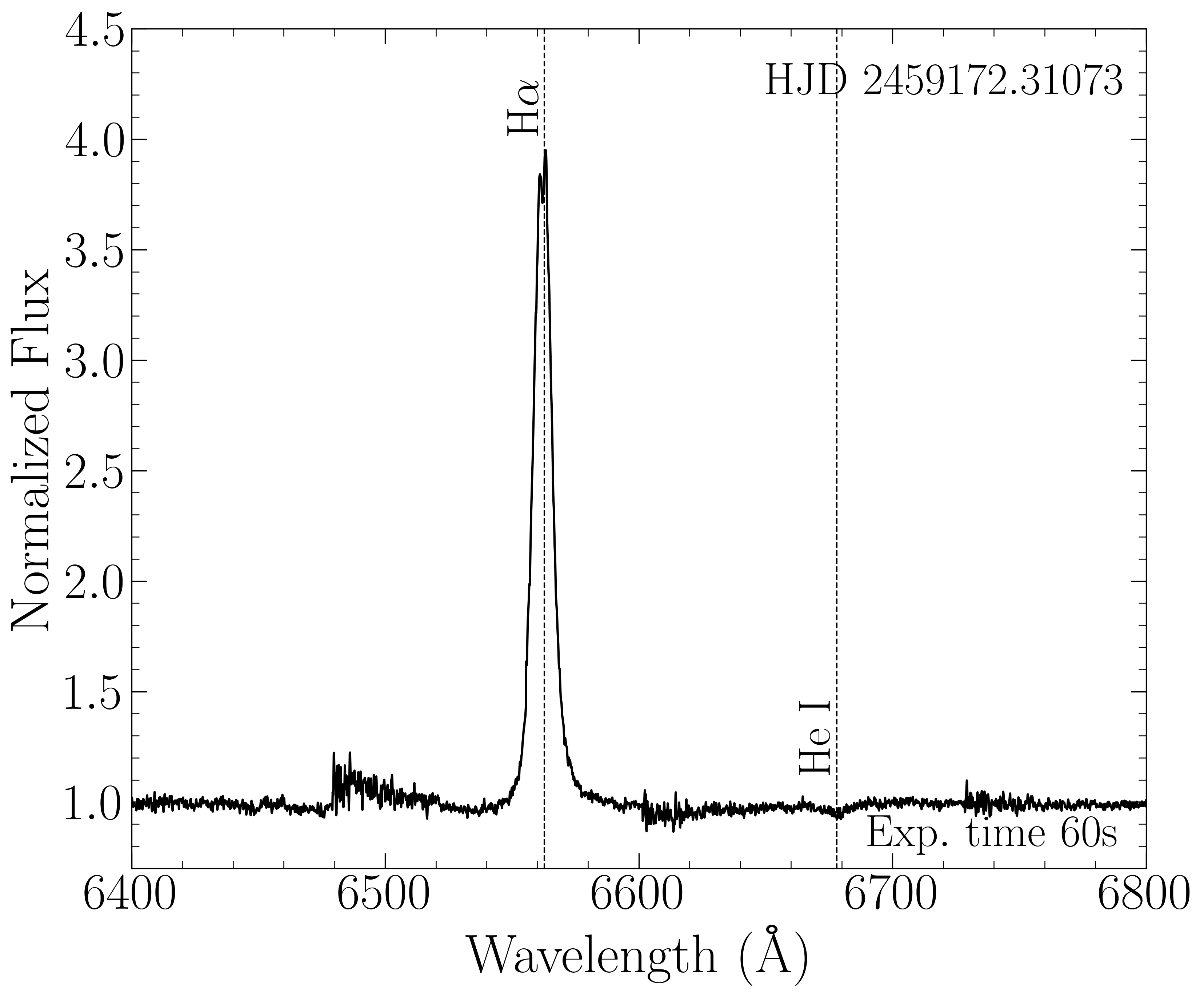}
         \caption{Two emission peaks in the H$\alpha$ region in the spectral data of EXO 051910+3737.7.}
         \label{fig:H-alphaspectrum}
     \end{subfigure}
     \hfill
     \begin{subfigure}[b]{0.48\textwidth}
         \centering
         \includegraphics[width=\textwidth]{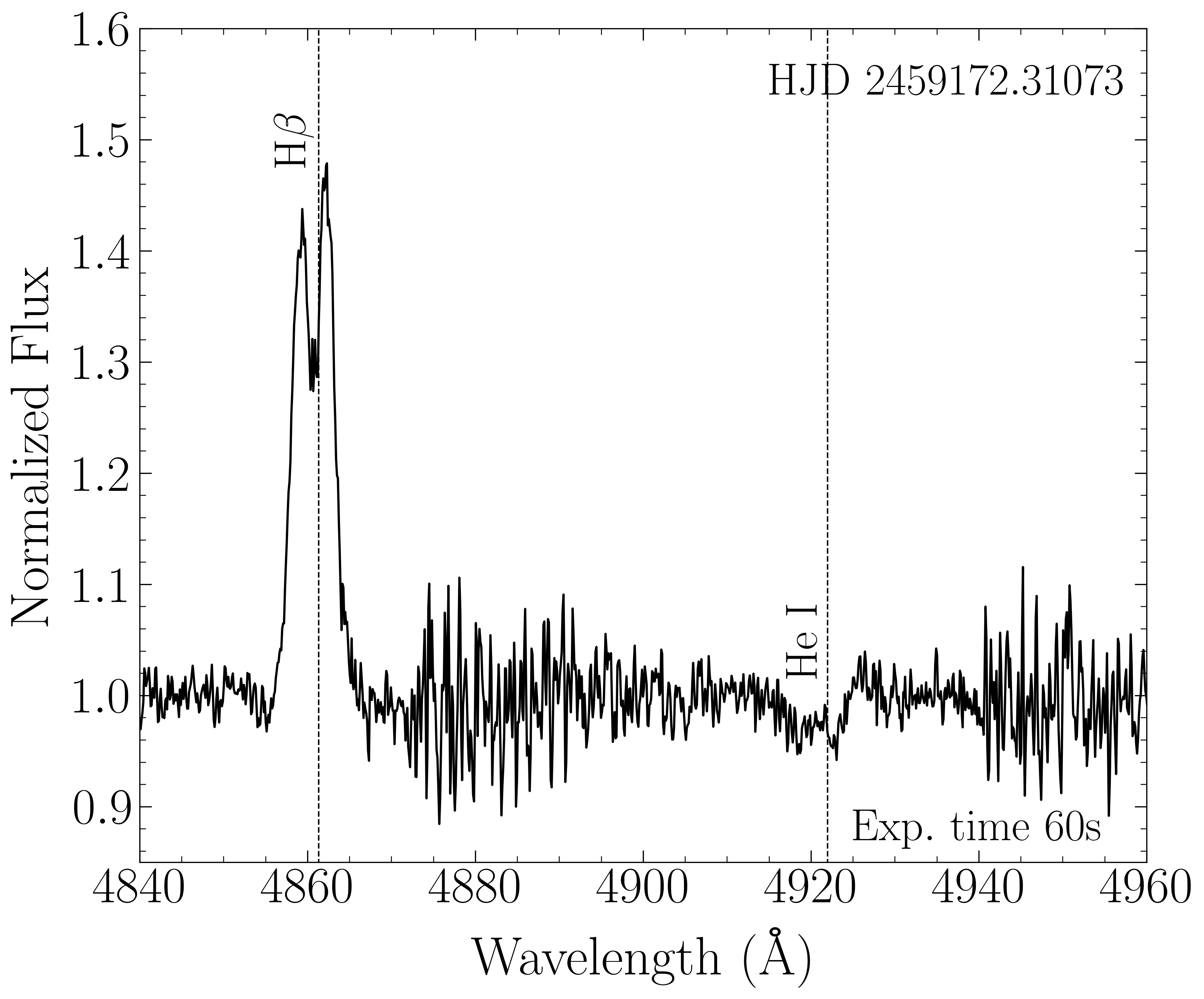}
         \caption{Two emission peaks in the H$\beta$ region in the spectral data of EXO 051910+3737.7.}
         \label{fig:H-betaspectrum}
     \end{subfigure}
        \caption{Samples of two-peak features of Balmer lines and He I region of EXO 051910+3737.7 were taken by TNO/MRES on 18 Nov 2020, on HJD 2459172.31073.}
        \label{fig:spectrum}
\end{figure}

In addition to Balmer lines, the spectra show clear He I lines in two wavelengths, He I $\lambda$6678 and He I $\lambda$4922. A weak He I $\lambda$6678 absorption is distinguishable above the noise in some observed spectra as shown in \textbf{Figure~\ref{fig:H-alphaspectrum}}. In the blue part, He I $\lambda$4922 shows the existing P Cygni profile in \textbf{Figure~\ref{fig:H-betaspectrum}}. 

Nevertheless, the metallic lines within the blue spectral region exhibit a relatively low S/N ratio, making it challenging to conduct a comprehensive analysis of these metallic lines.

\section*{\normalsize Line profile fitting methodology}

After detecting a change in the Balmer line peak, we proceeded to study its physical properties. During the fitting, we found that only the H$\beta$ line can be analyzed because H$\alpha$ line has an amplitude four times as high as defined as the delta function, and it is quite complex to find the peaks. Therefore, we will study the optical thickness properties of circumstellar discs from H$\beta$ line.

The Gaussian model was used for line fitting to characterize H$\beta$ lines. Given the observed splitting in the Hydrogen Balmer line, notably evident in H$\beta$, we applied the doubled-Gaussian profile to distinguish between two central peaks. The least mean square method was employed to achieve the best-fit lines. The general form of its Gaussian probability density function is

\begin{equation}
G(x) = \frac{1}{{\sigma \sqrt {2\pi } }}e^{{{ - \left( {x - \mu } \right)^2 } \mathord{\left/ {\vphantom {{ - \left( {x - \mu } \right)^2 } {2\sigma ^2 }}} \right. \kern-\nulldelimiterspace} {2\sigma ^2 }}},
\end{equation}
where $\mu$ refers to the mean, and $\sigma$ is standard deviation.

We employed the \texttt{lmfit} package along with its associated function to extract specific parameters from the spectra lines, specifically, the mean, standard deviation, amplitude, full-width at half-maximum (FWHM), height of the spectra, and chi-square values ($\chi^2$). The fitted results are shown in \textbf{Figure~\ref{Fig:gaussionfitting}} which H$\alpha$ line emission spectra (solid lines) along with the theoretical line profiles (dotted lines) for different Be stars. The $\chi^2$ values are reported in \textbf{Table~\ref{Table:VR}}. When determining initial parameters, it is crucial to observe the data closely. Upon visual inspection, it becomes evident that the two peaks of the Gaussian function display a discernible separation, although there is also a degree of overlap between them.

\begin{figure}[htbp]
\begin{subfigure}[t]{0.29\textwidth}
    \includegraphics[width=\linewidth]{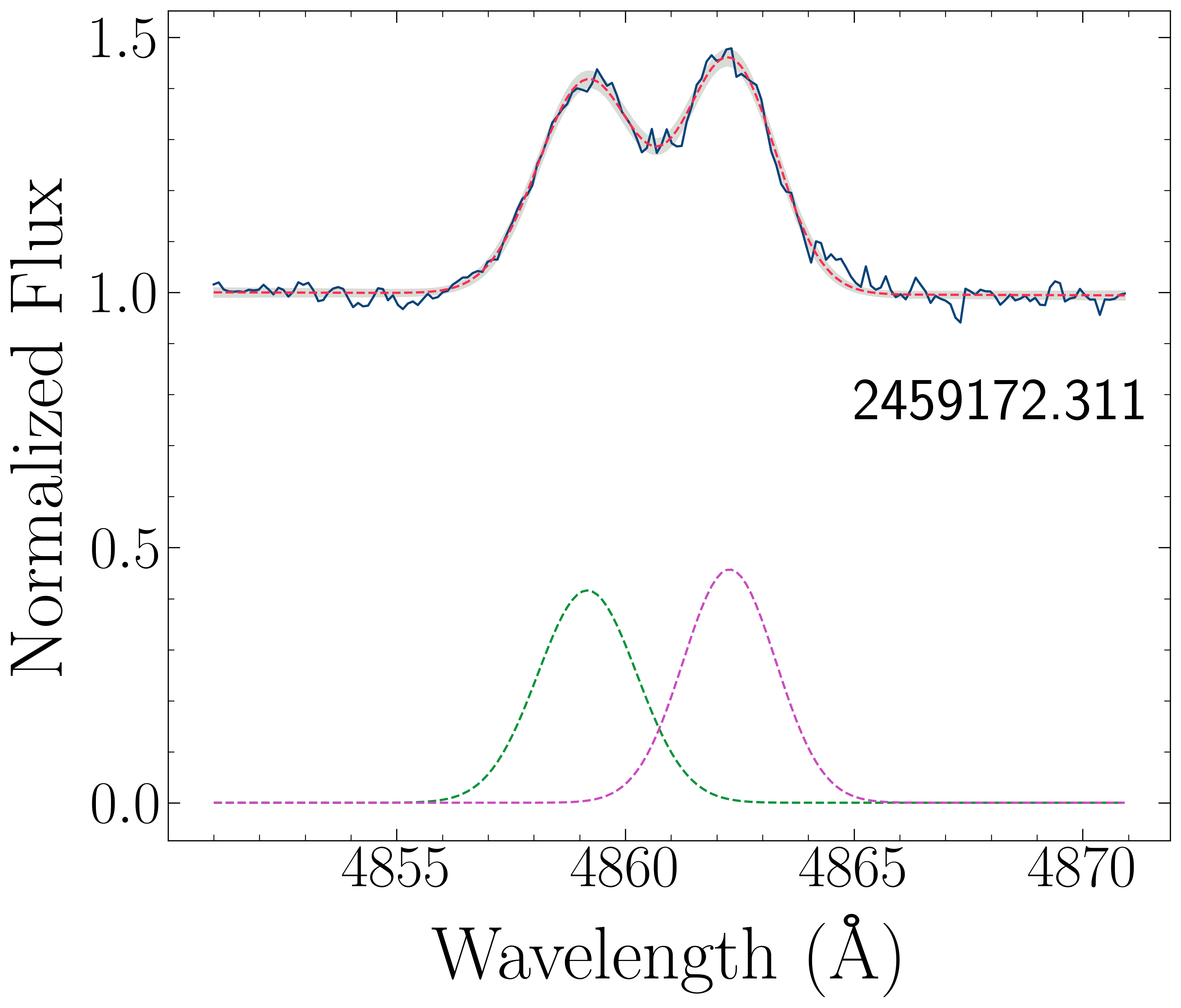}

\end{subfigure}\hfill
\begin{subfigure}[t]{0.29\textwidth}
  \includegraphics[width=\linewidth]{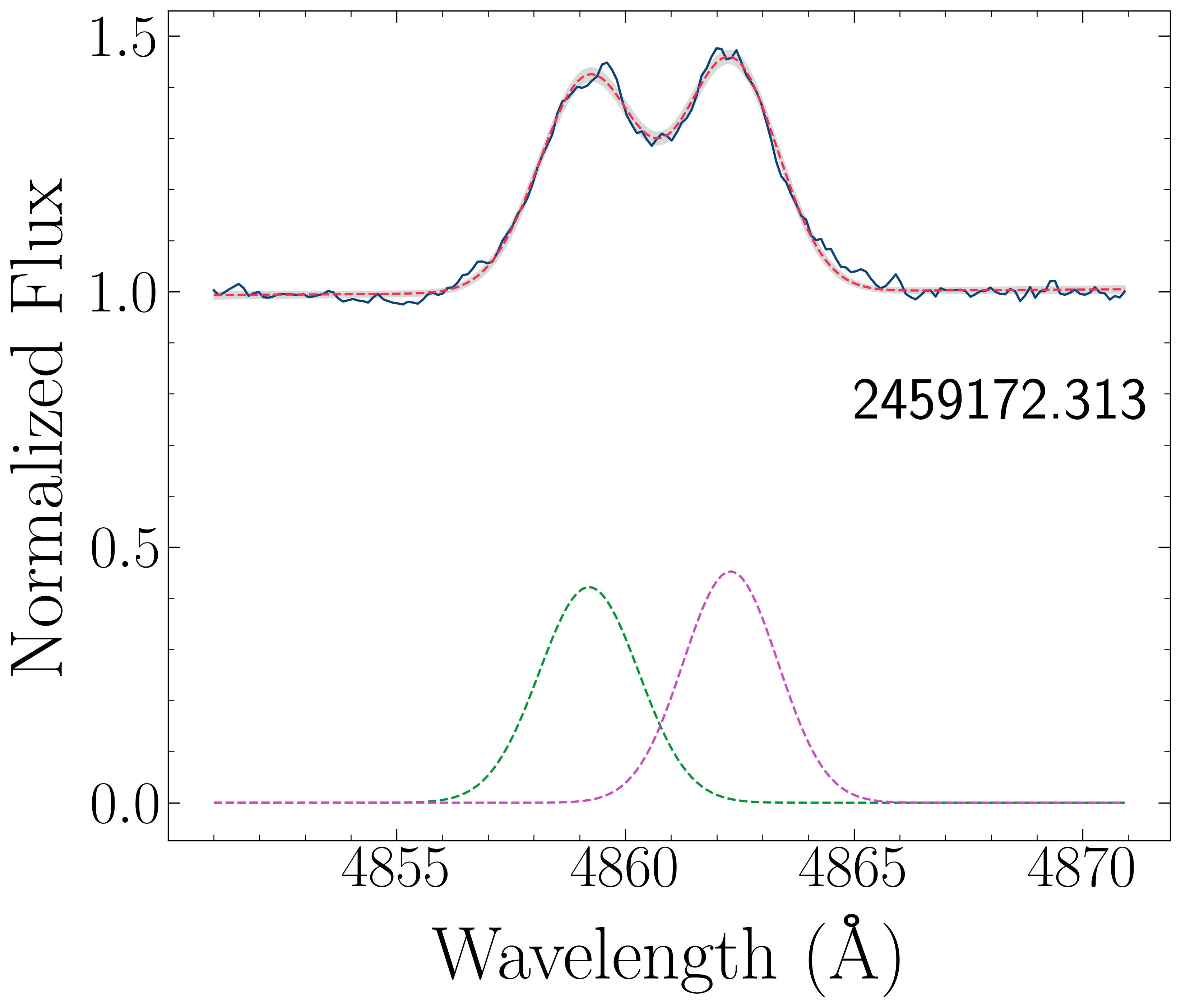}

\end{subfigure}\hfill
\begin{subfigure}[t]{0.29\textwidth}
    \includegraphics[width=\linewidth]{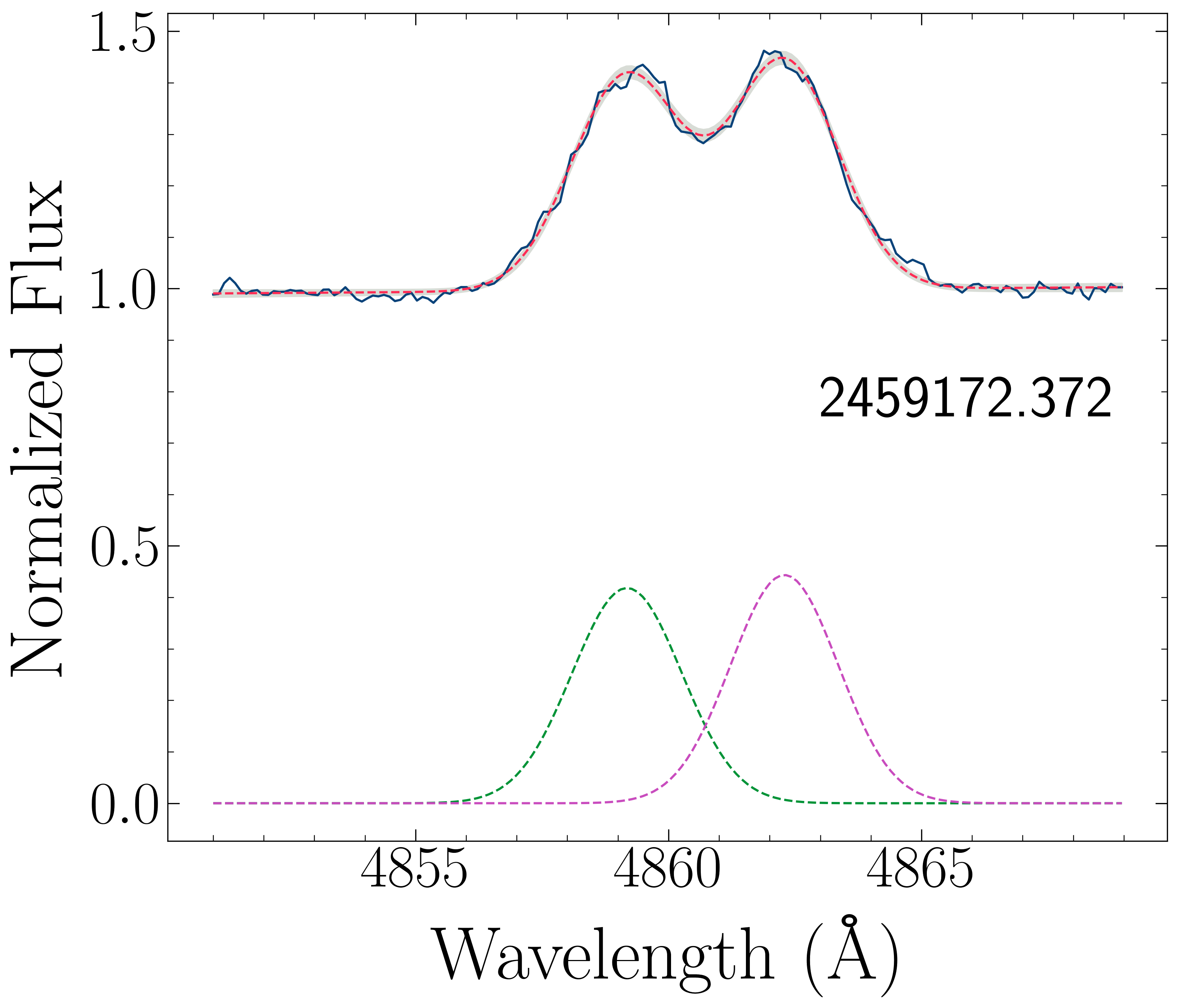}

\end{subfigure}

\begin{subfigure}[t]{0.29\textwidth}
    \includegraphics[width=\linewidth]{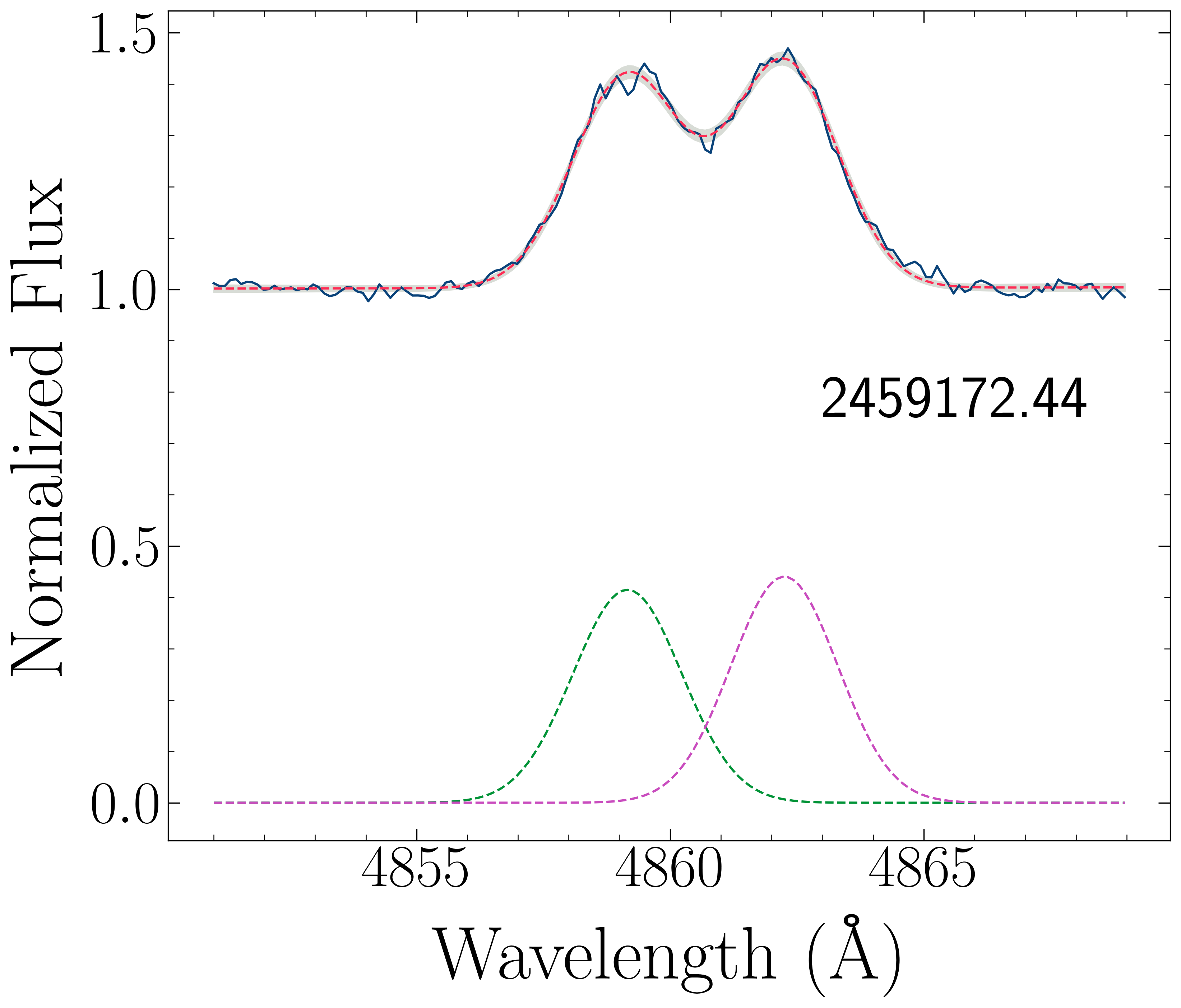}

\end{subfigure}\hfill
\begin{subfigure}[t]{0.29\textwidth}
    \includegraphics[width=\linewidth]{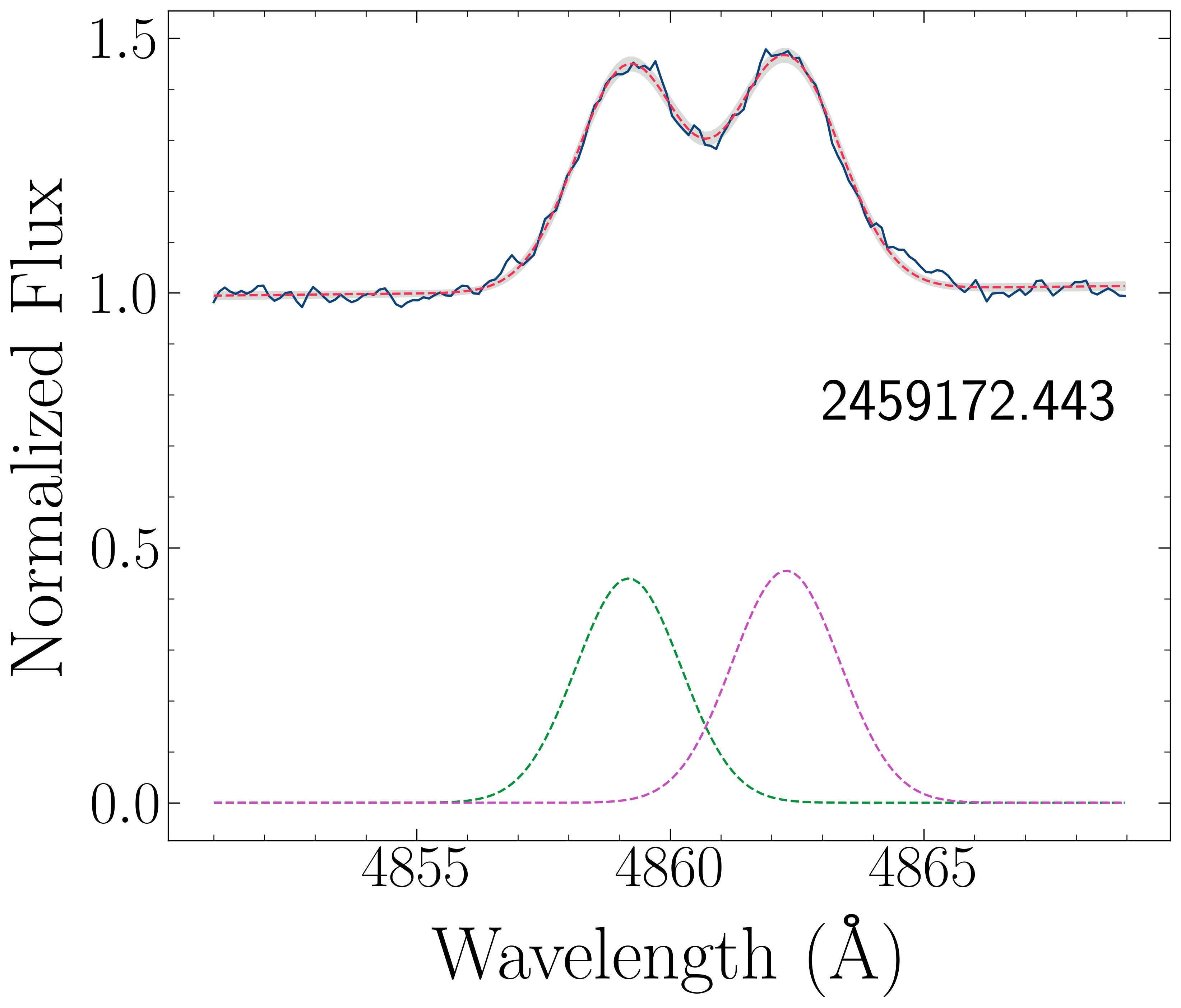}

\end{subfigure}\hfill
\begin{subfigure}[t]{0.29\textwidth}
    \includegraphics[width=\textwidth]{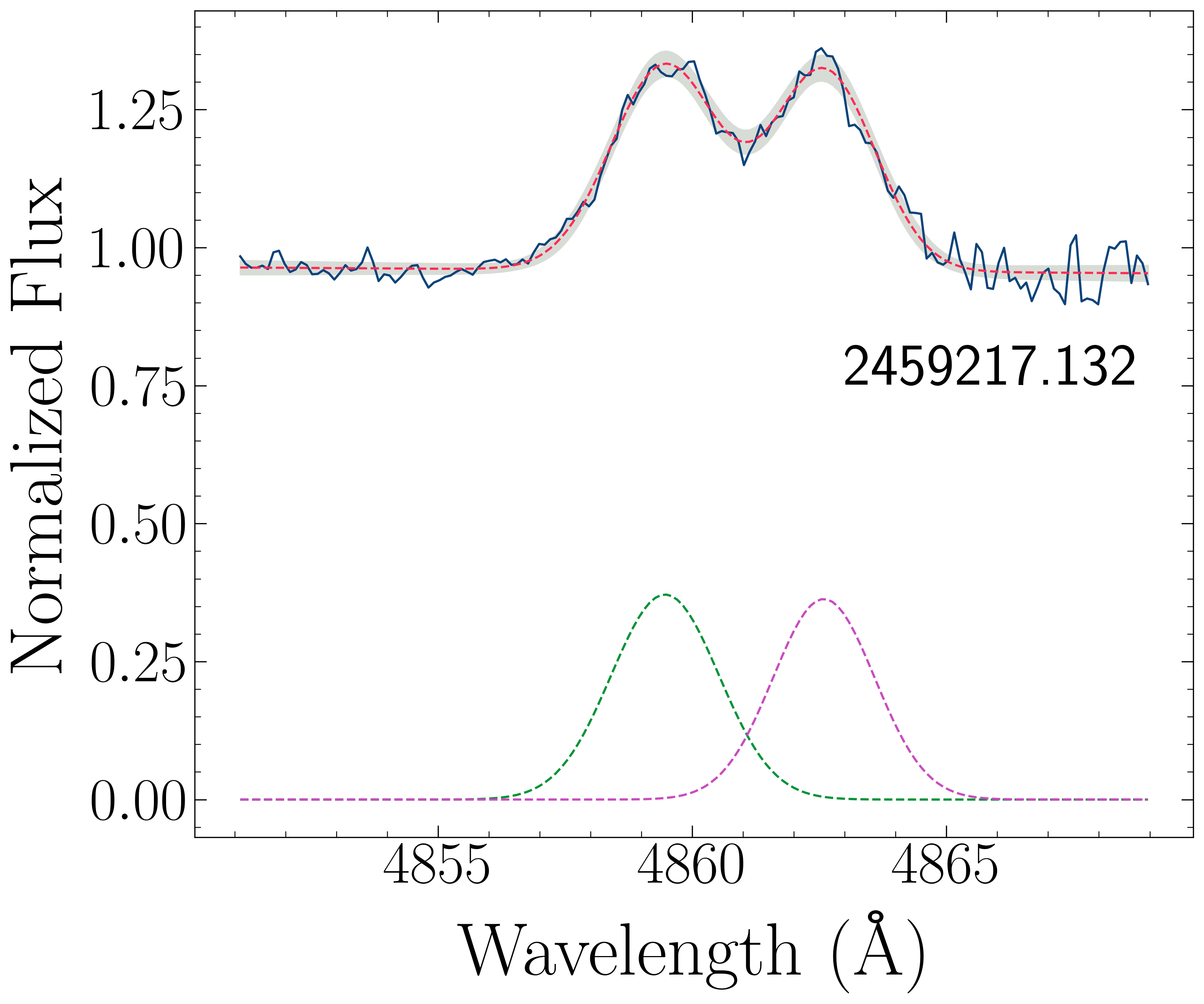}

\end{subfigure}

\caption{The schematic of all observed H$\beta$ spectral lines (blue solid lines) along with the best-fit lines (red dotted lines) using a doubled Gaussian model implemented through the \texttt{lmfit} package in Python, employing the least mean square method. The 22 spectral observations were gathered over seven nights. Additionally, the green and purple lines represent the individual Gaussian components. The shaded grey area illustrates the 3-$\sigma$ uncertainty band.}
\label{Fig:gaussionfitting}
\end{figure}

\renewcommand{\thefigure}{\arabic{figure} }
\addtocounter{figure}{-1}

\begin{figure}[htbp]
\begin{subfigure}[t]{0.29\textwidth}
    \includegraphics[width=\linewidth]{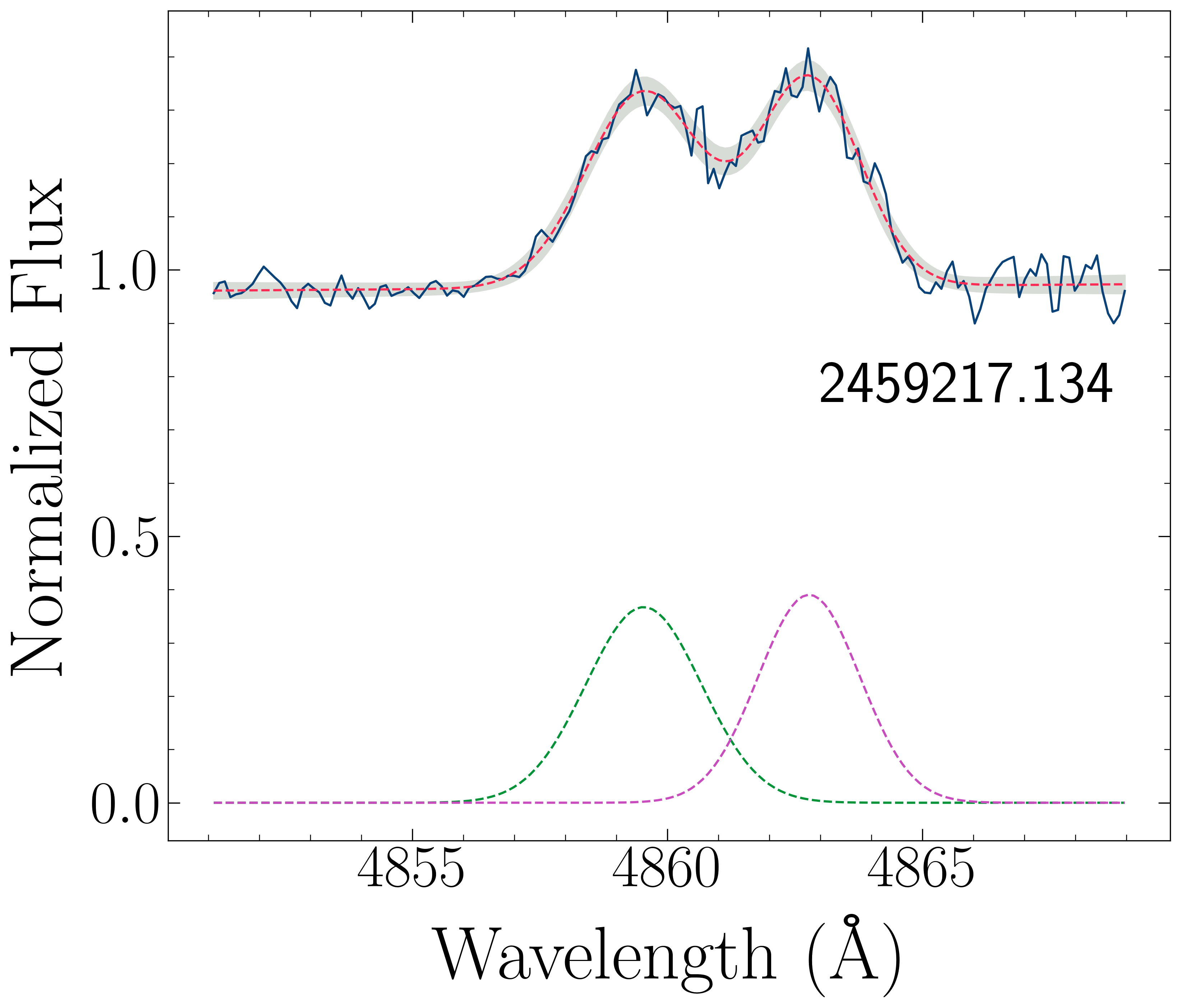}

\end{subfigure}\hfill
\begin{subfigure}[t]{0.29\textwidth}
  \includegraphics[width=\linewidth]{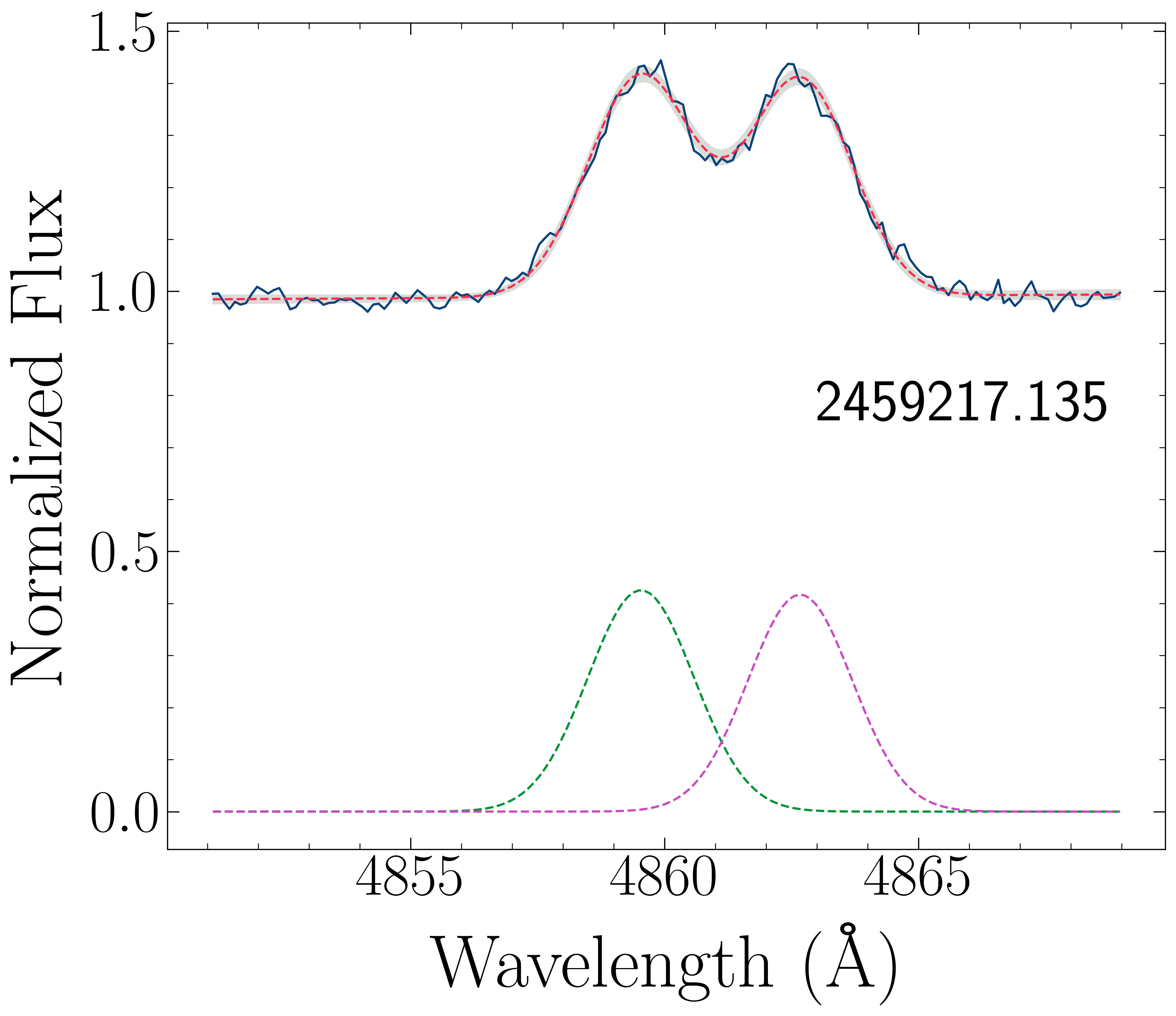}

\end{subfigure}\hfill
\begin{subfigure}[t]{0.29\textwidth}
    \includegraphics[width=\linewidth]{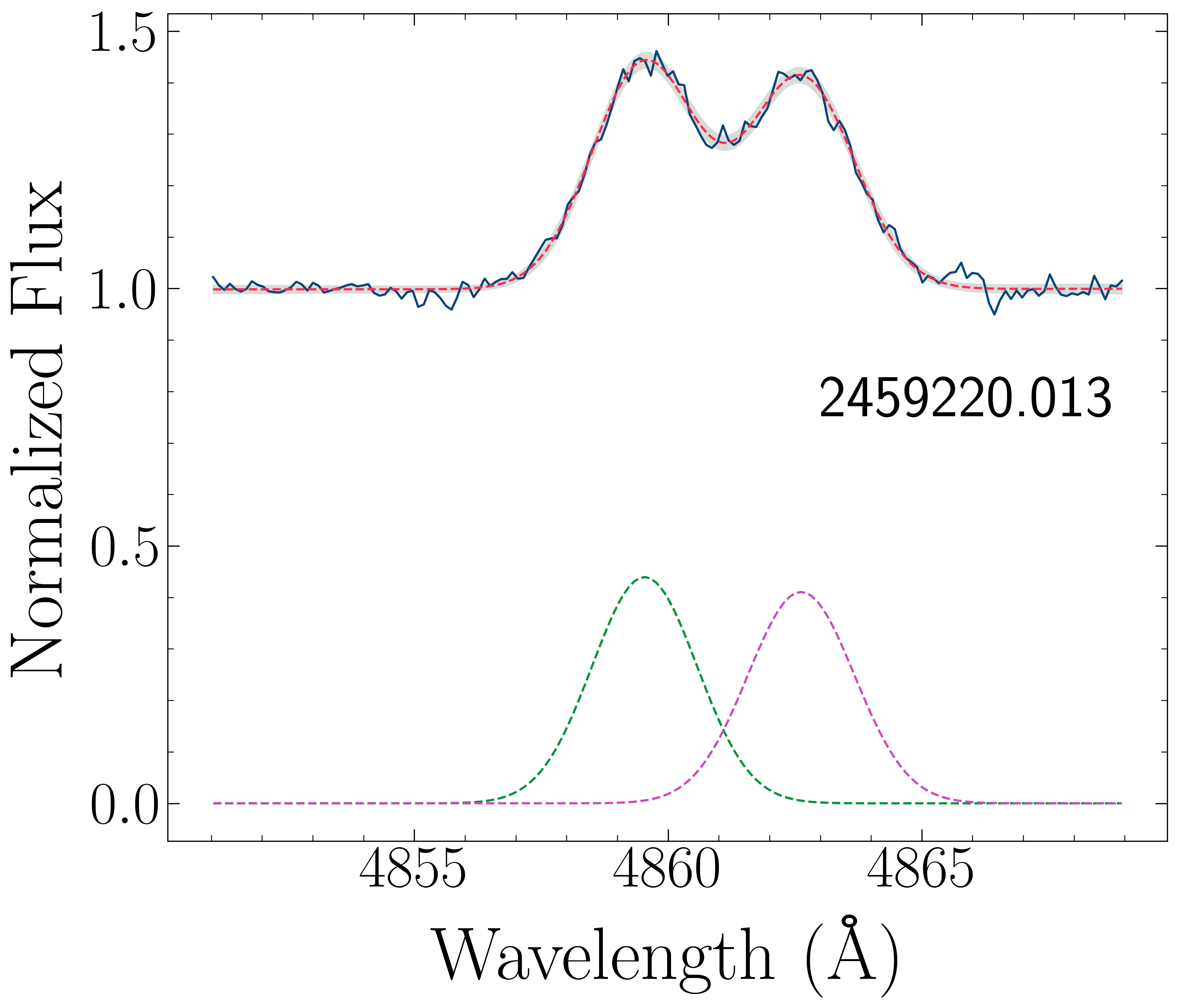}

\end{subfigure}

\begin{subfigure}[t]{0.29\textwidth}
    \includegraphics[width=\linewidth]{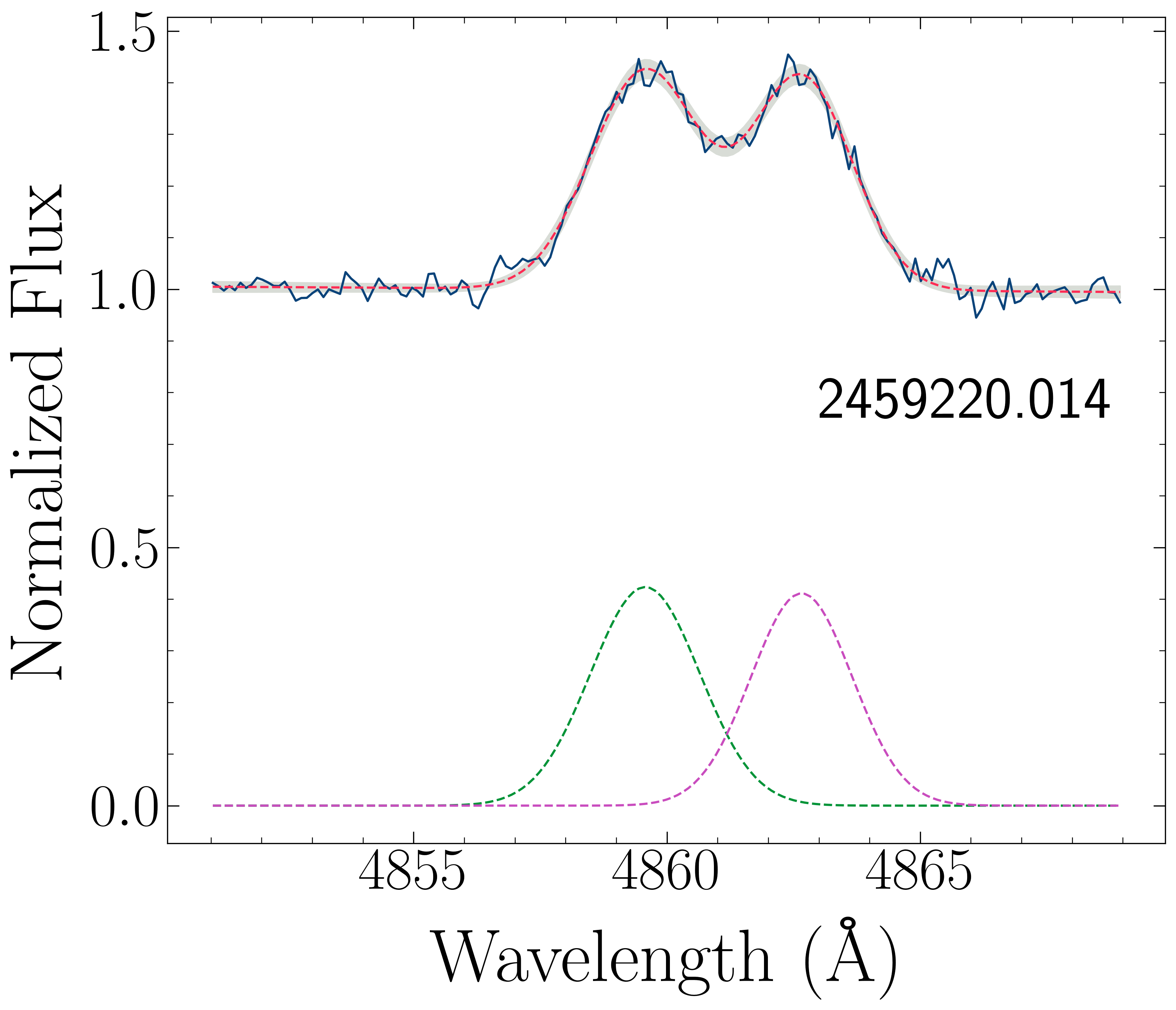}

\end{subfigure}\hfill
\begin{subfigure}[t]{0.29\textwidth}
    \includegraphics[width=\linewidth]{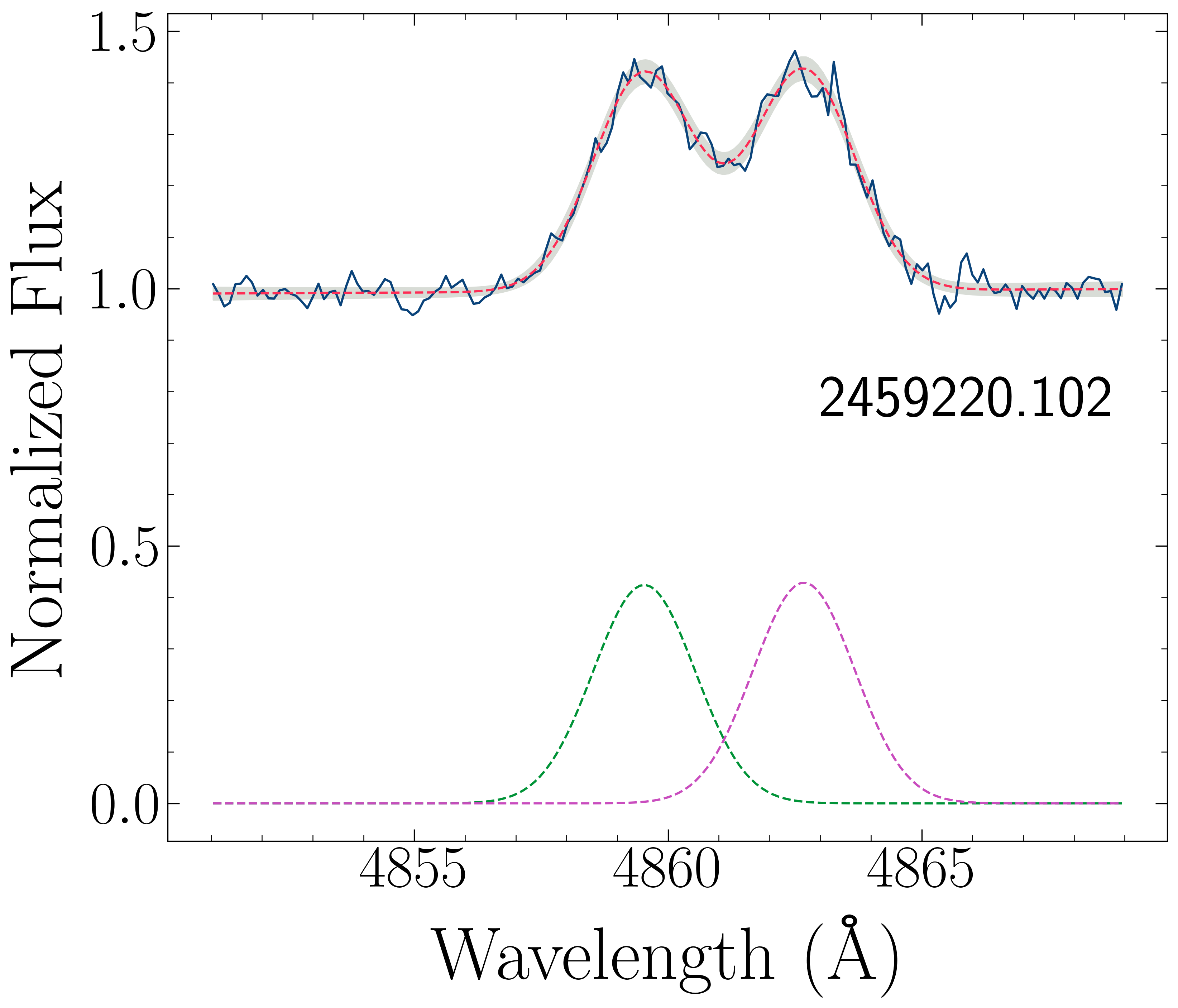}

\end{subfigure}\hfill
\begin{subfigure}[t]{0.29\textwidth}
    \includegraphics[width=\textwidth]{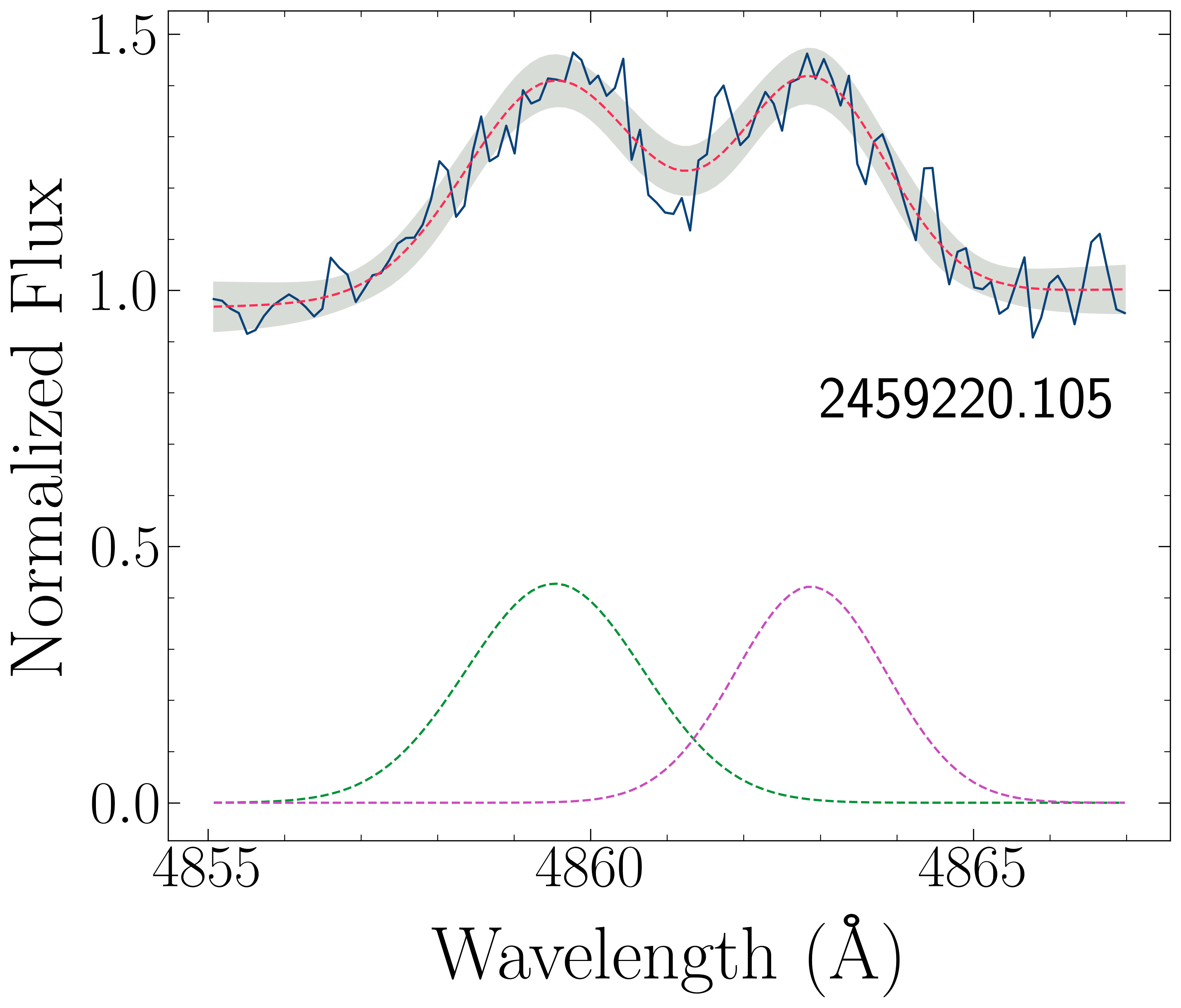}
\end{subfigure}

\begin{subfigure}[t]{0.29\textwidth}
    \includegraphics[width=\linewidth]{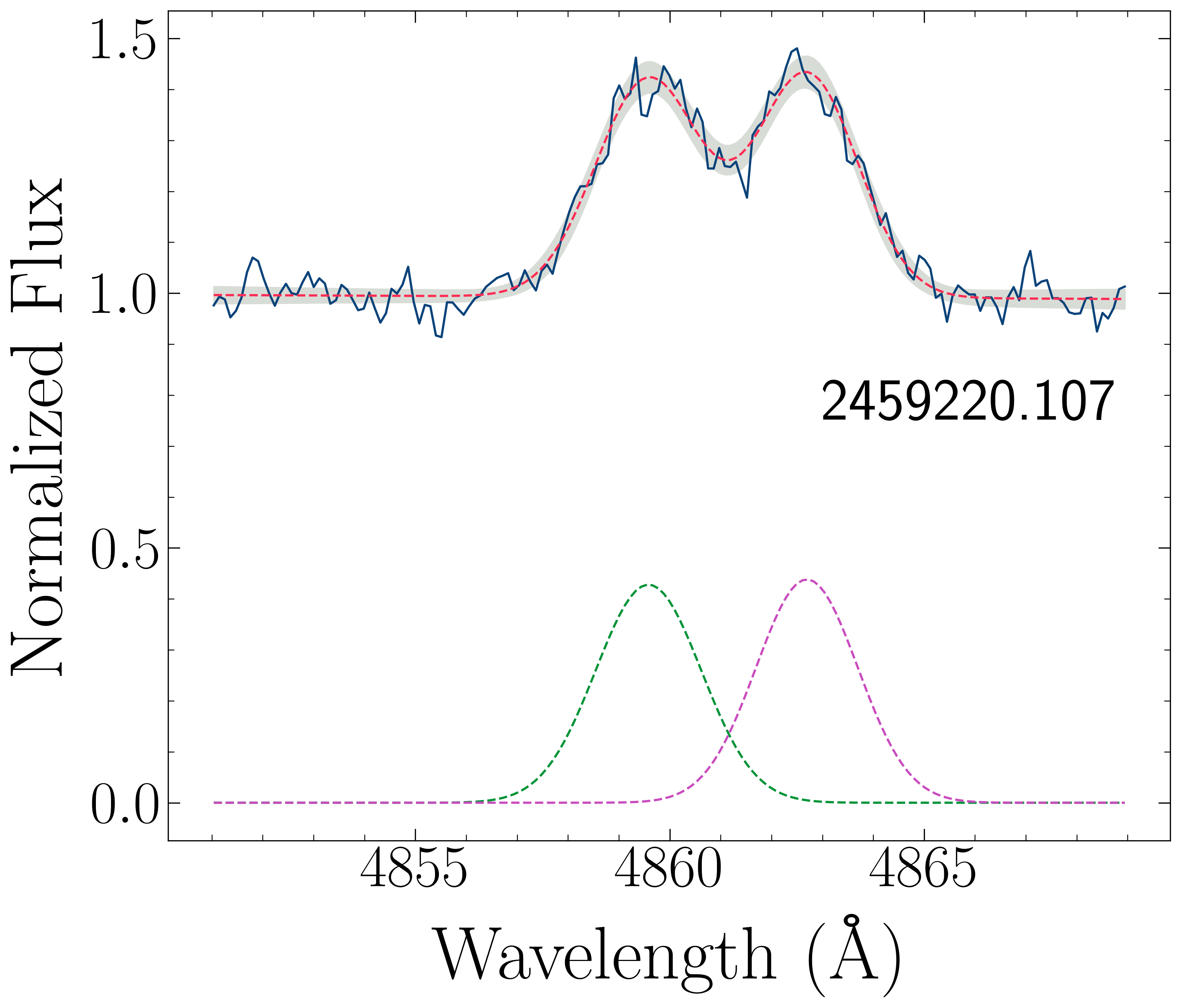}

\end{subfigure}\hfill
\begin{subfigure}[t]{0.29\textwidth}
  \includegraphics[width=\linewidth]{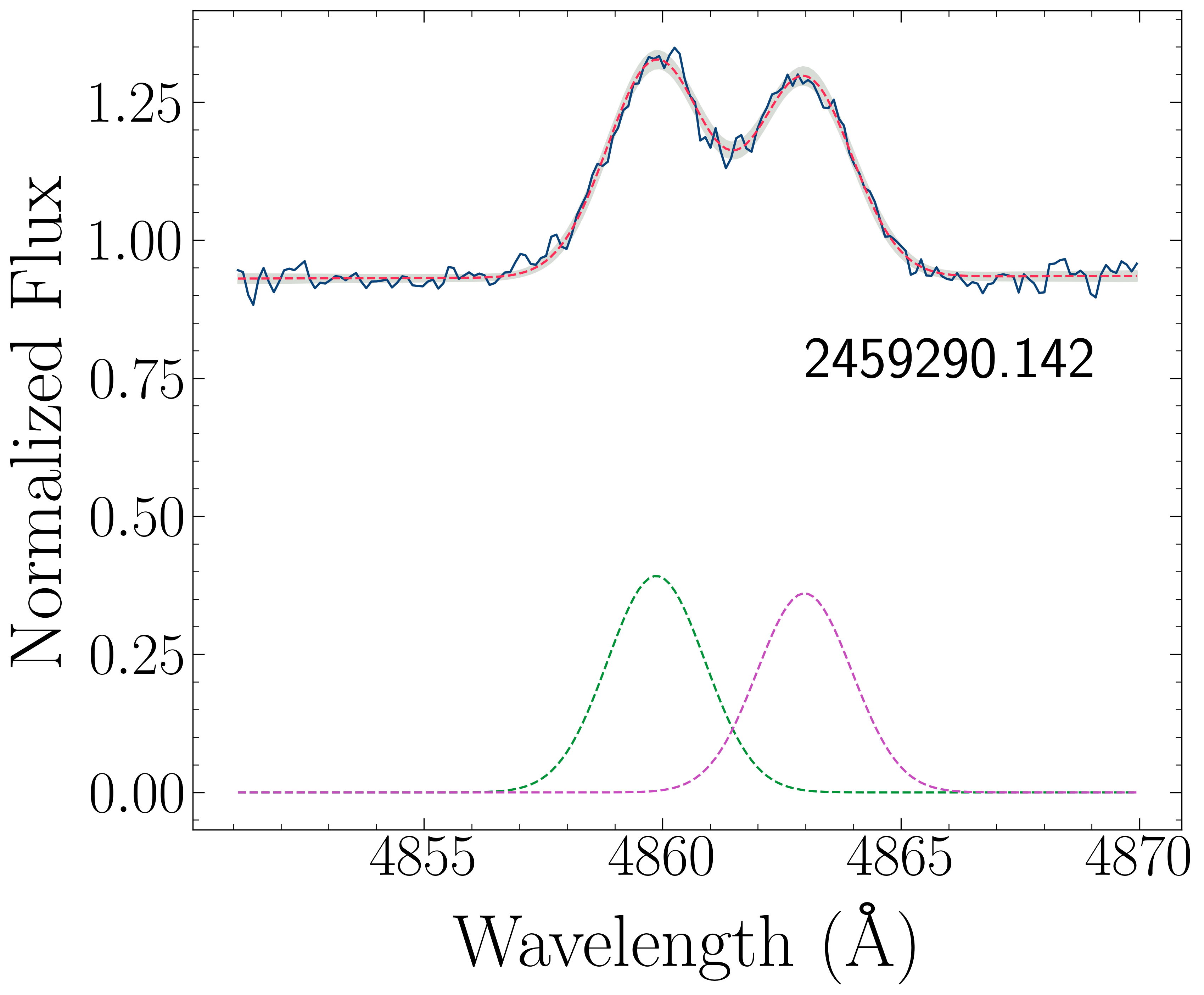}

\end{subfigure}\hfill
\begin{subfigure}[t]{0.29\textwidth}
    \includegraphics[width=\linewidth]{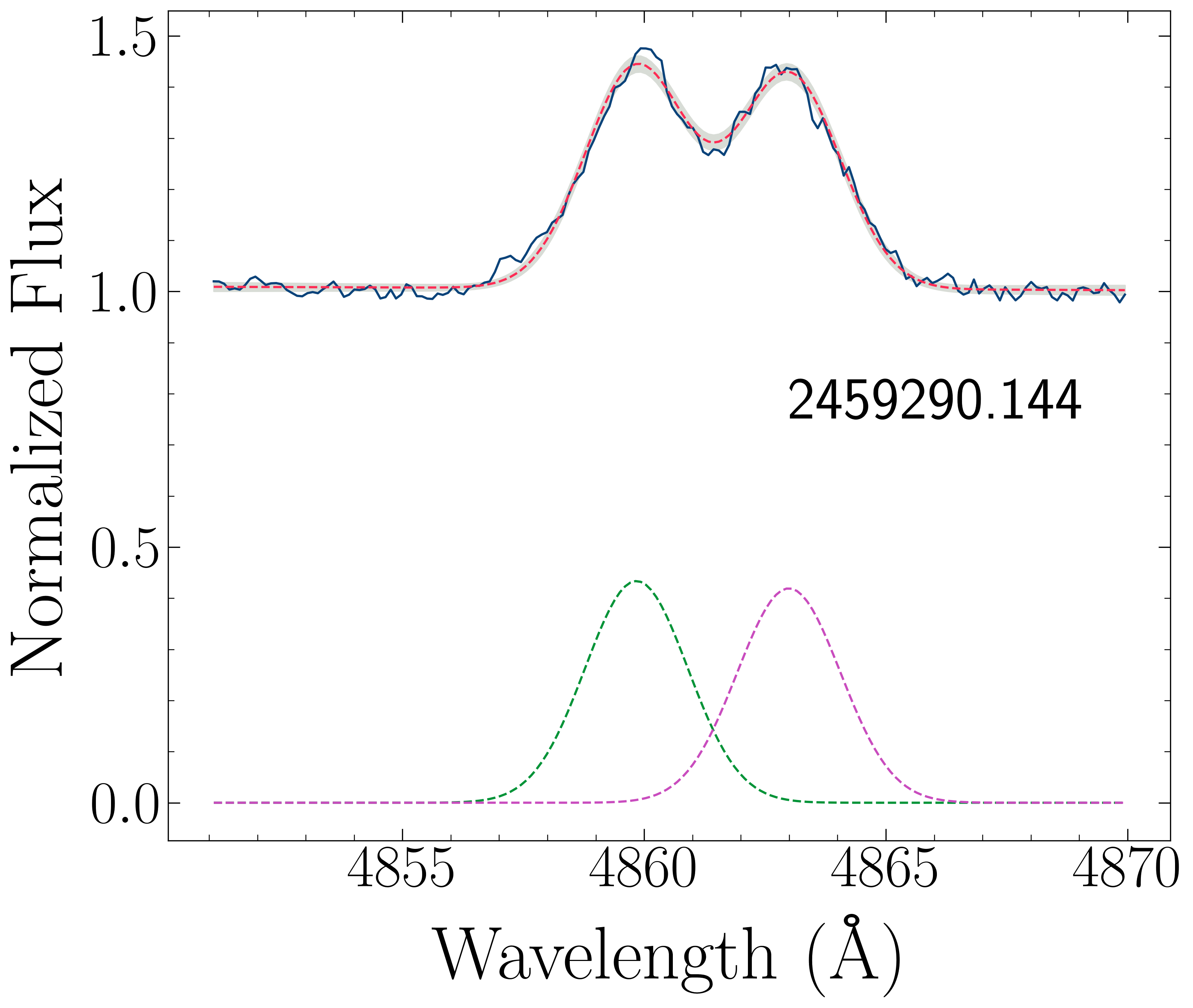}

\end{subfigure}

\begin{subfigure}[t]{0.29\textwidth}
    \includegraphics[width=\linewidth]{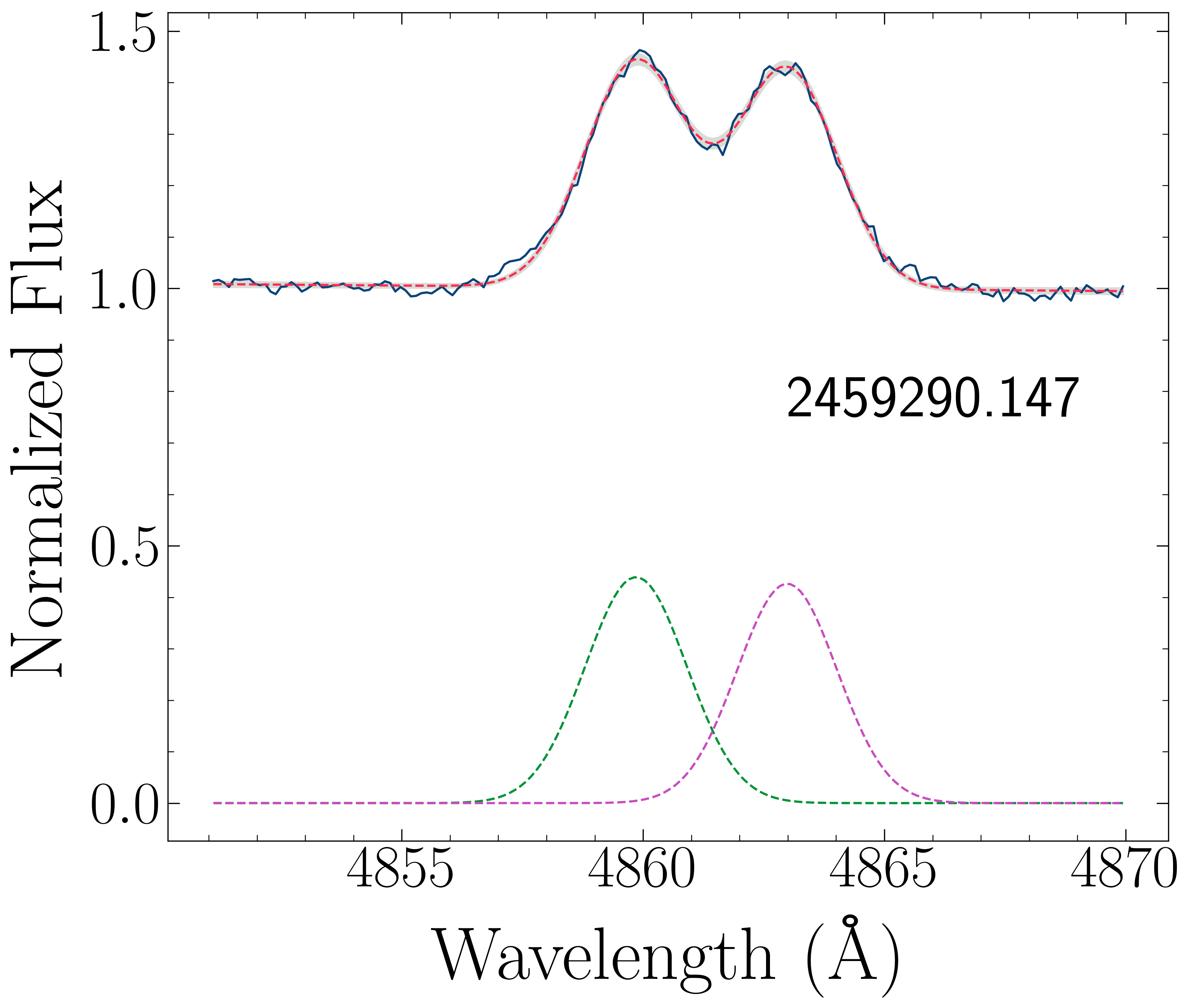}

\end{subfigure}\hfill
\begin{subfigure}[t]{0.29\textwidth}
    \includegraphics[width=\linewidth]{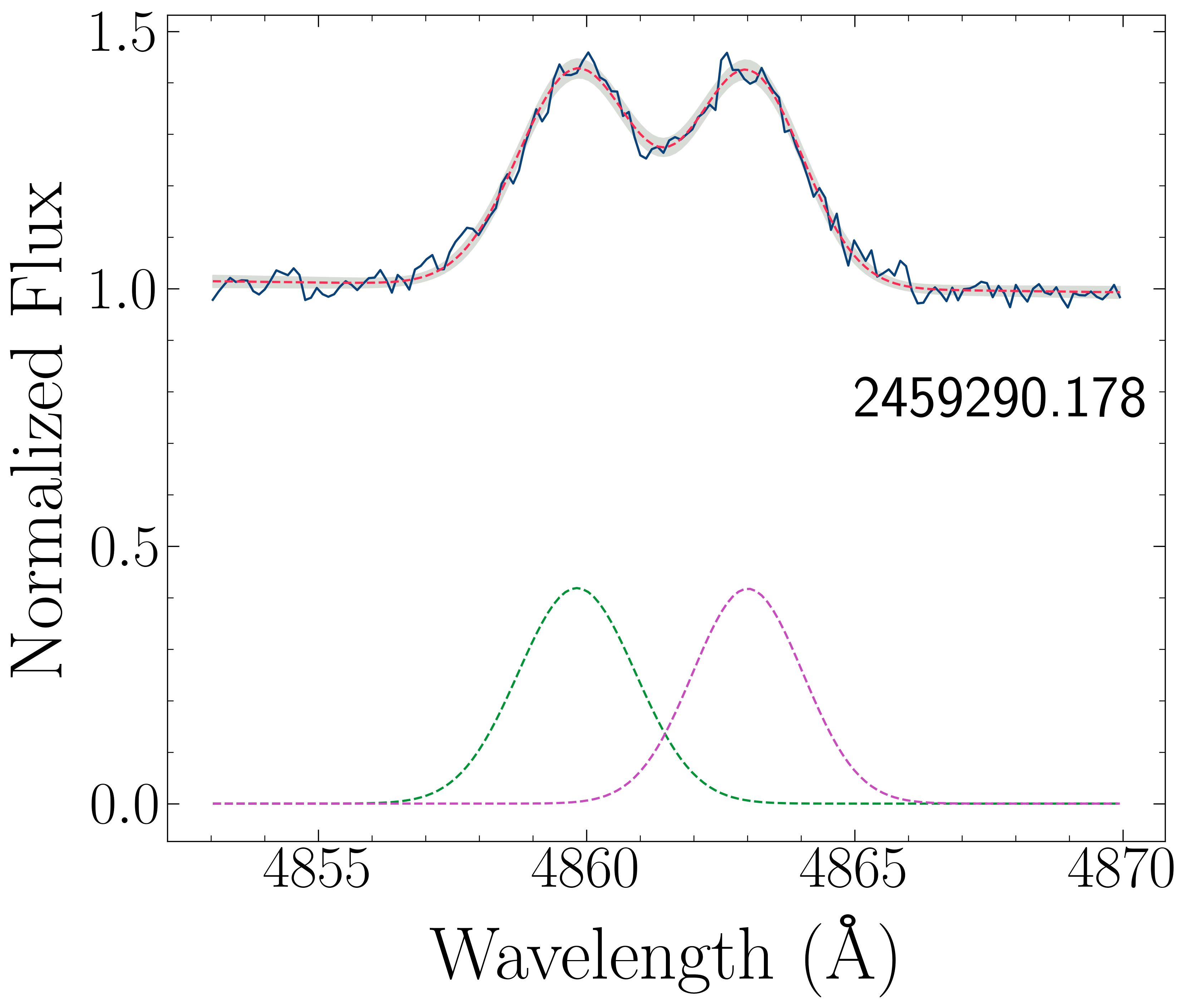}

\end{subfigure}\hfill
\begin{subfigure}[t]{0.29\textwidth}
    \includegraphics[width=\textwidth]{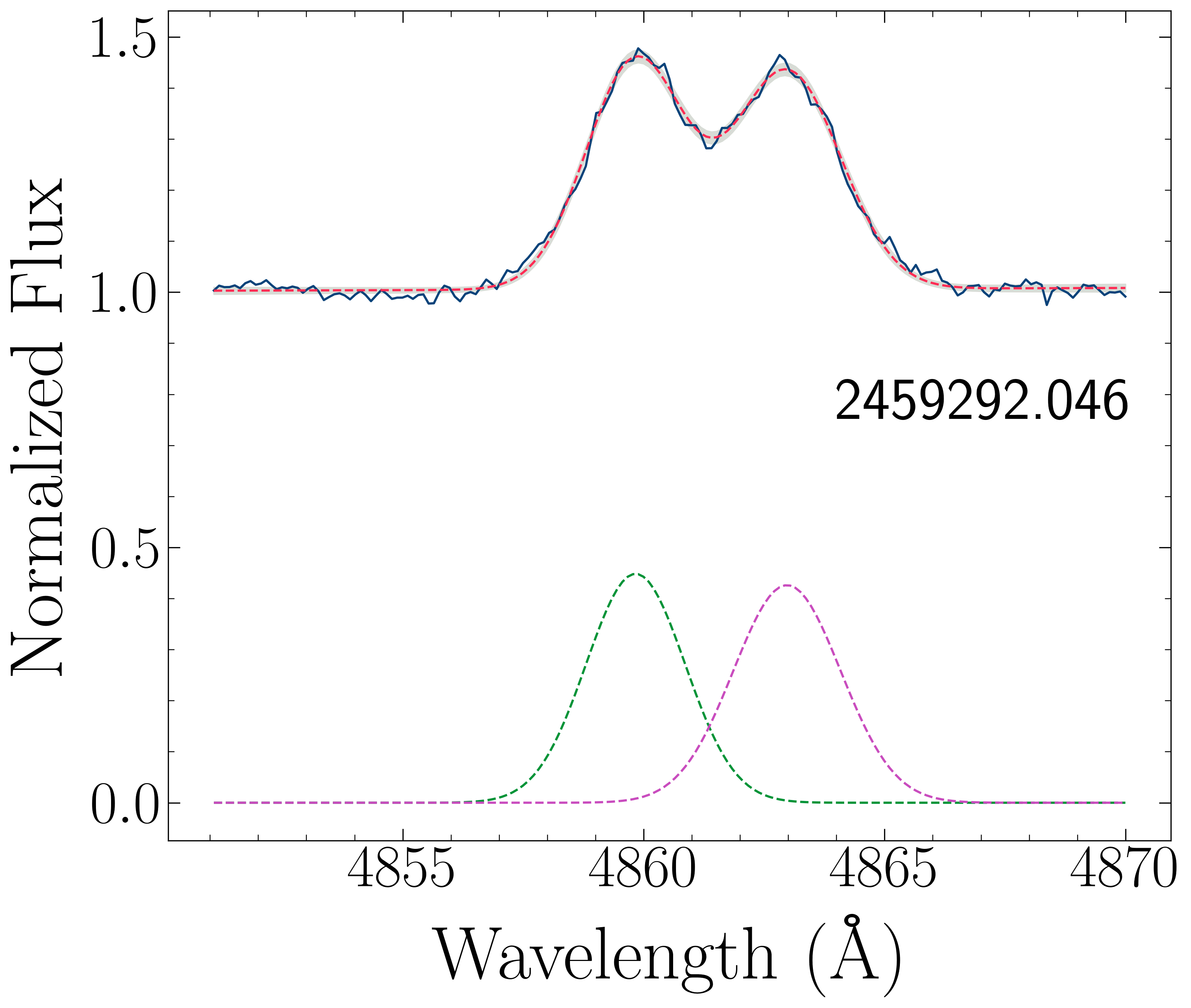}
\end{subfigure}

\begin{subfigure}[t]{0.29\textwidth}
    \includegraphics[width=\linewidth]{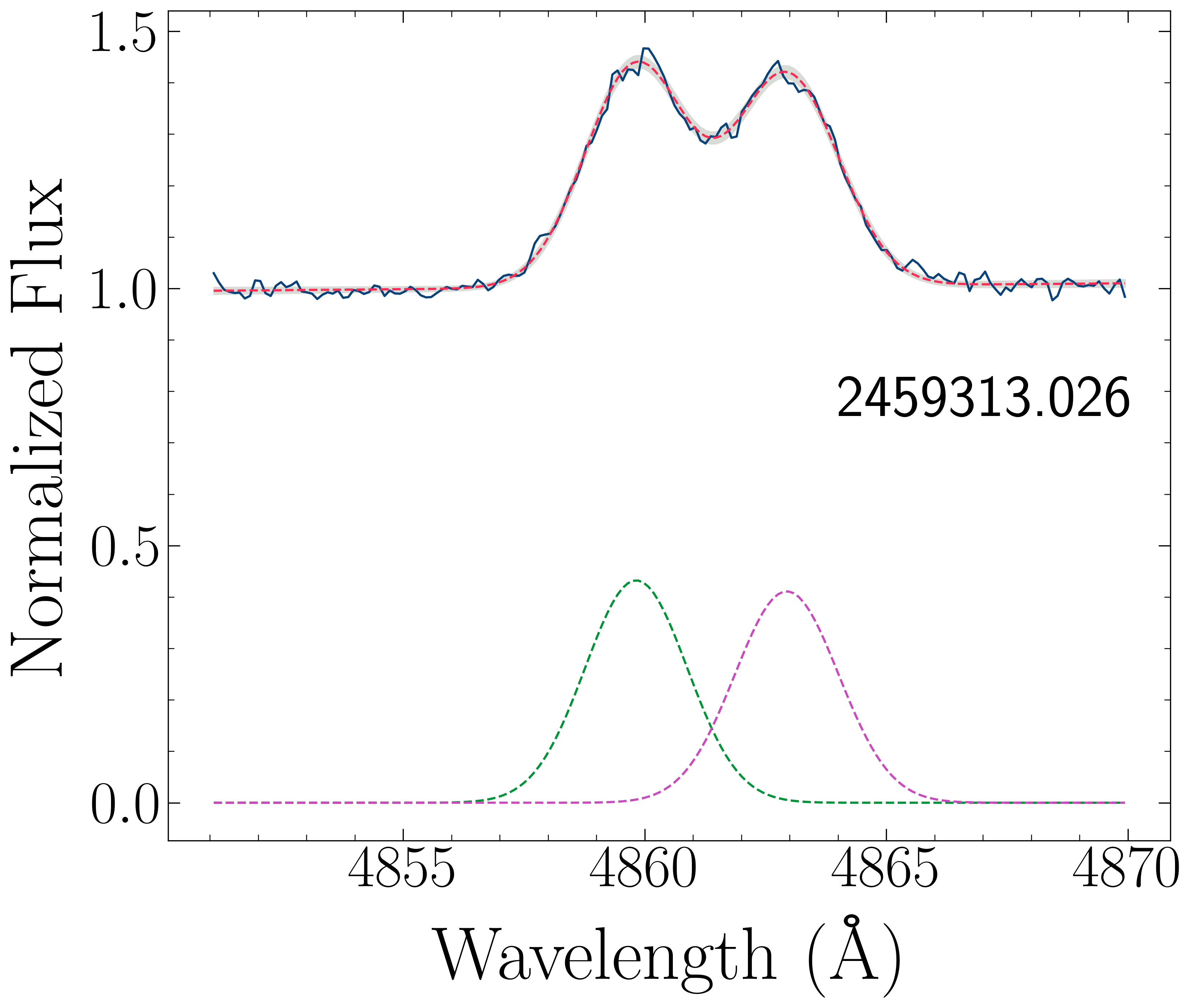}

\end{subfigure}\hfill
\begin{subfigure}[t]{0.29\textwidth}
    \includegraphics[width=\linewidth]{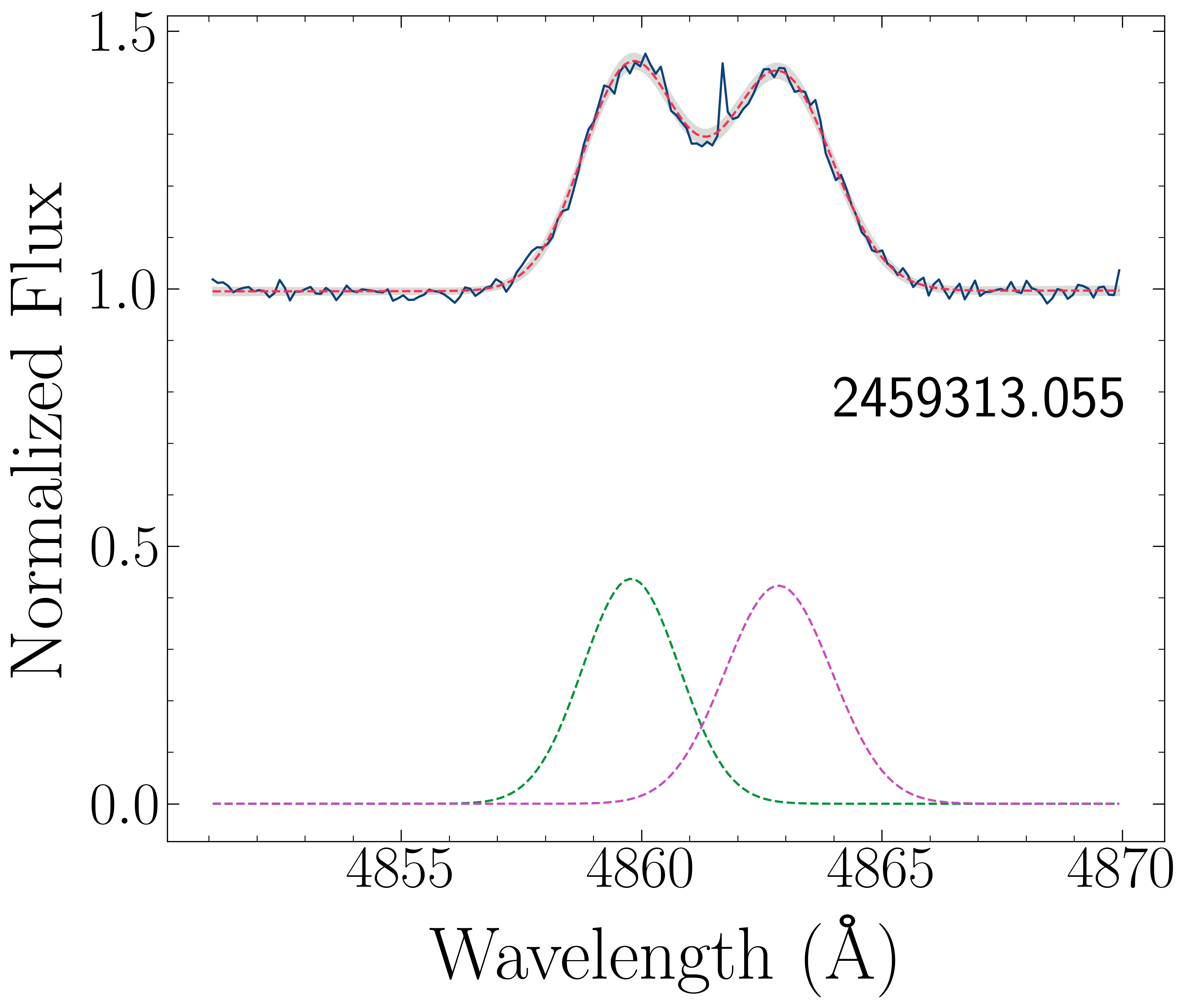}

\end{subfigure}\hfill
\begin{subfigure}[t]{0.29\textwidth}
  \includegraphics[width=\linewidth]{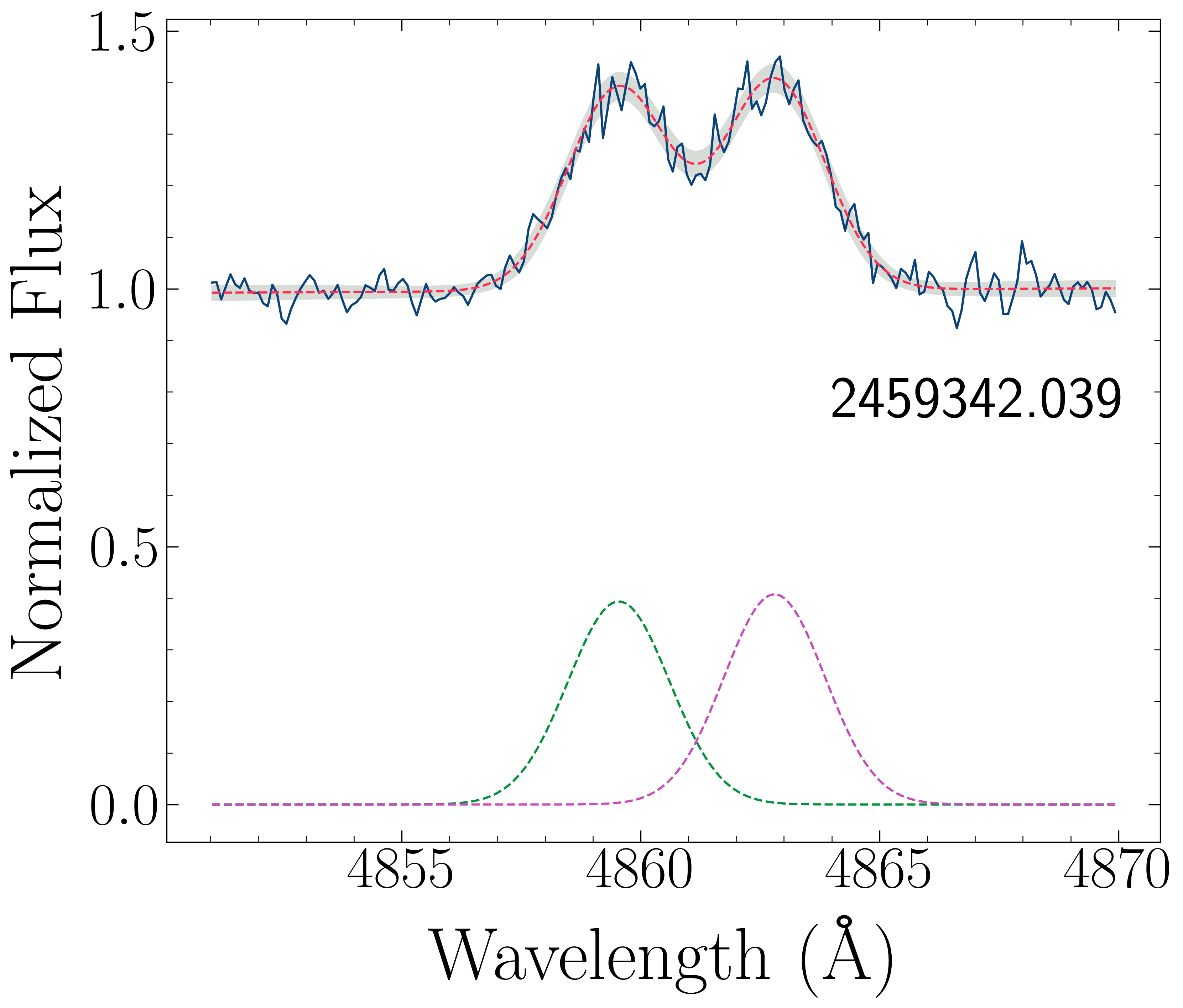}

\end{subfigure}
\begin{subfigure}[t]{0.29\textwidth}
    \includegraphics[width=\linewidth]{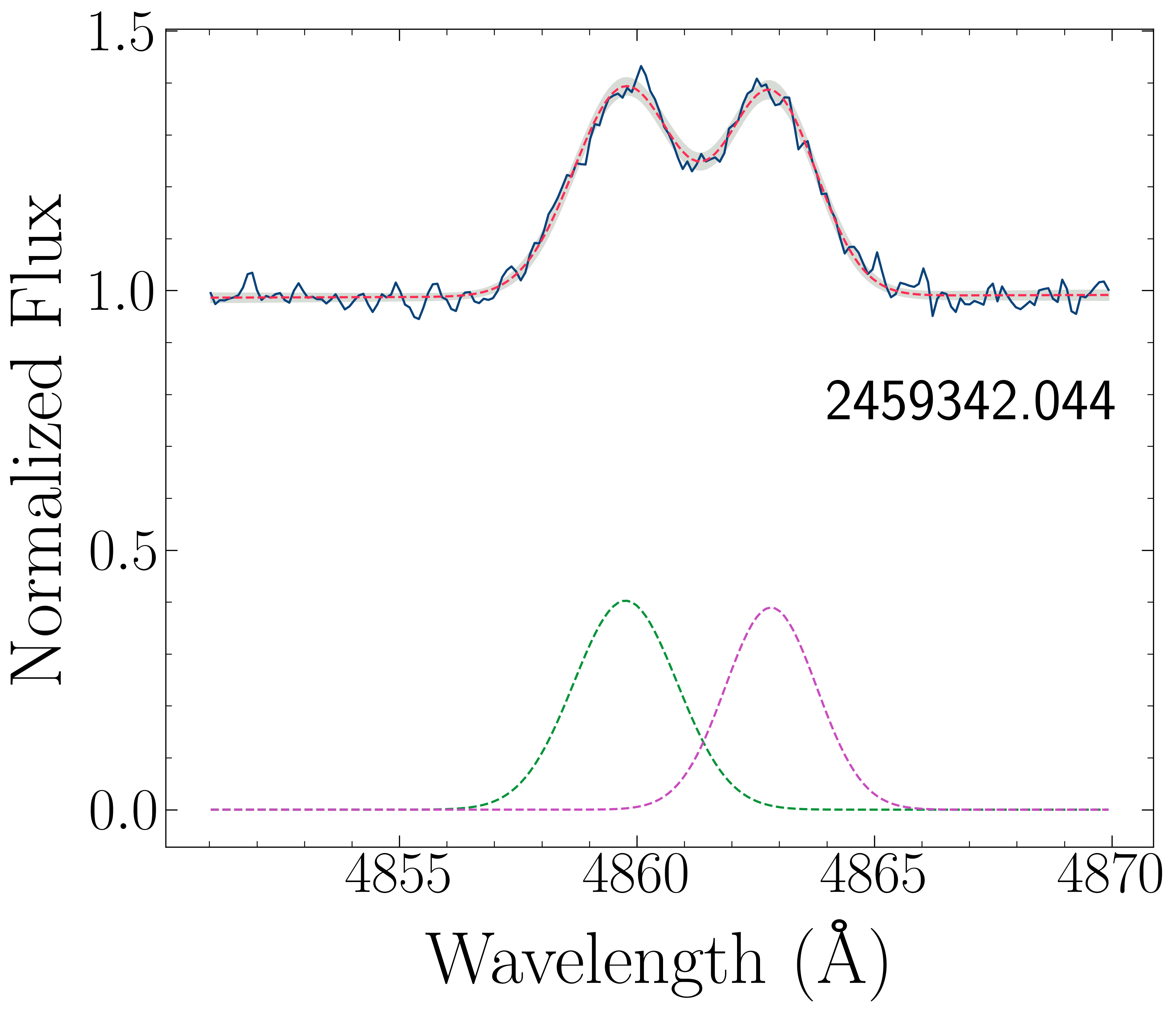}

\end{subfigure}

\caption{\textit{--continued}}
\end{figure}

\renewcommand{\thefigure}{\arabic{figure}}



\section*{\normalsize V/R variation} 
As mentioned in the previous section, H$\alpha$ and H$\beta$ lines mainly refer to the emission lines. This spectral profile is perfectly normal for a Be star. We can see the double-peaked profiles in emission lines of V and R components in both H$\alpha$ and H$\beta$ lines. The V/R ratio was calculated from the V and R spectral peak intensities, which are variable on a timescale. V/R variability predominantly occurs during active phases and follows quasi-periodic cycles, characterized by the sequence V>R $\rightarrow$ V=R $\rightarrow$ V<R $\rightarrow$ V=R. Previous studies of V/R variations of Be stars indicate the presence of shorter-term variations occurring over days and months \cite{dachs1986}. It is noteworthy that V/R variability is independent of spectral type within the range of B0 to B5 stars \cite{hirata1981}. 

\begin{figure}[!h]
\centering
\includegraphics[width=\textwidth]{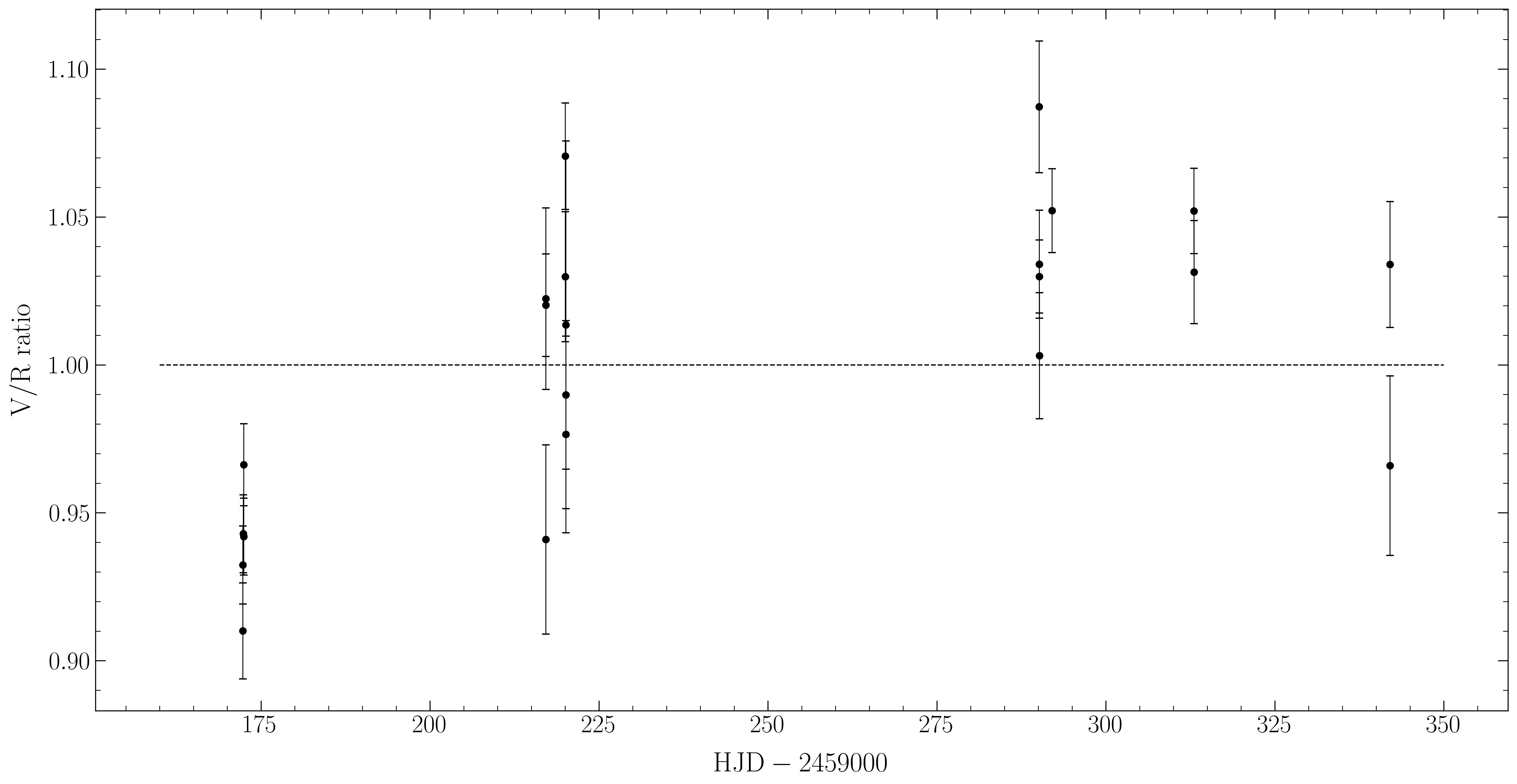}
\caption{Variations in the V/R ratio as a function of time for the H$\beta$ line with associated uncertainties over seven nights between 2020-2021.}
\label{Fig:vr_variation}
\end{figure}

\begin{figure}
     \centering
     \begin{subfigure}[b]{\textwidth}
         \centering
         \includegraphics[width=0.9\textwidth]{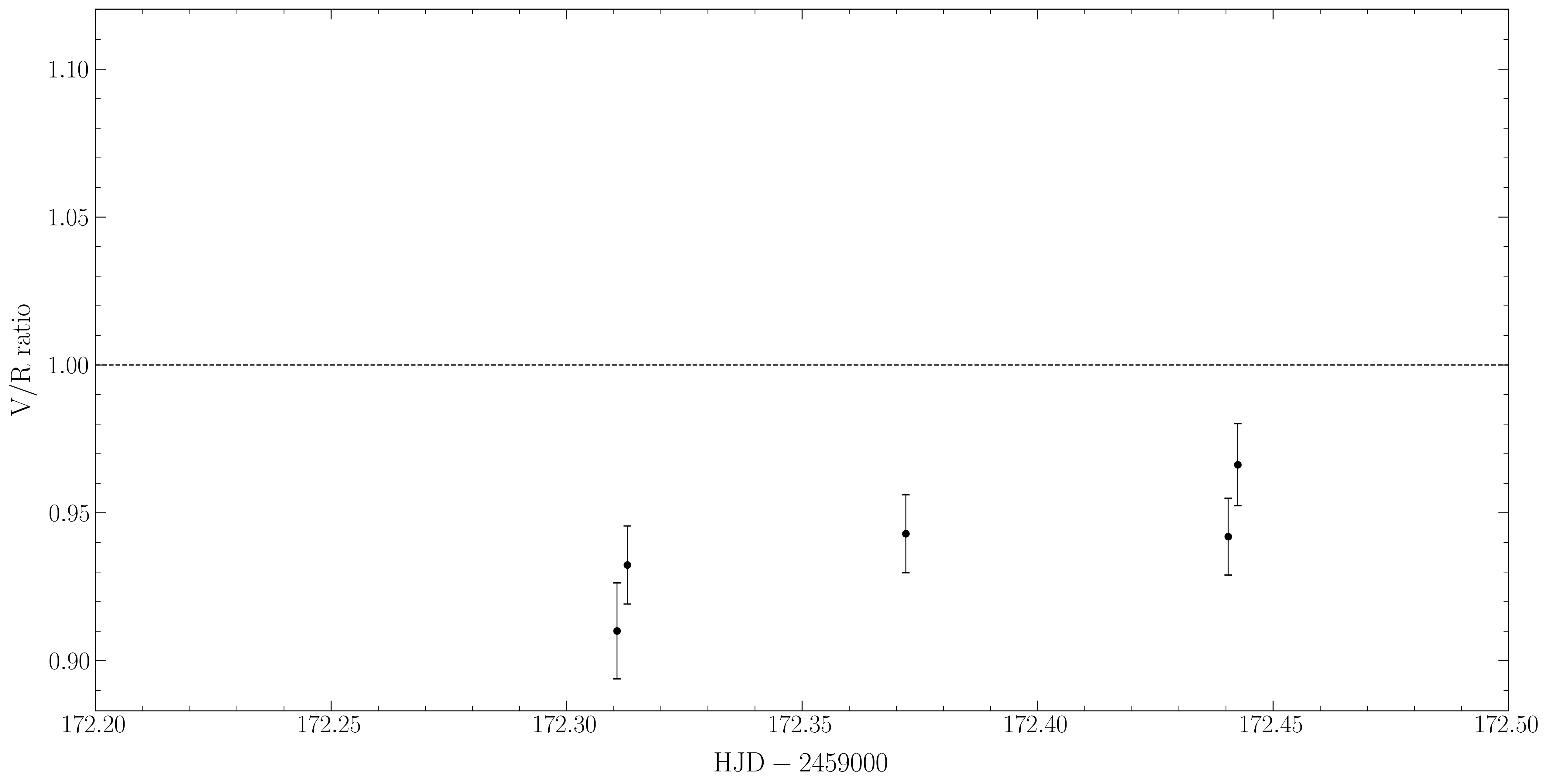}
         \caption{V/R variation in HJD 2459172.}
         \label{fig:172}
     \end{subfigure}
     \hfill
     \begin{subfigure}[b]{\textwidth}
         \centering
         \includegraphics[width=0.9\textwidth]{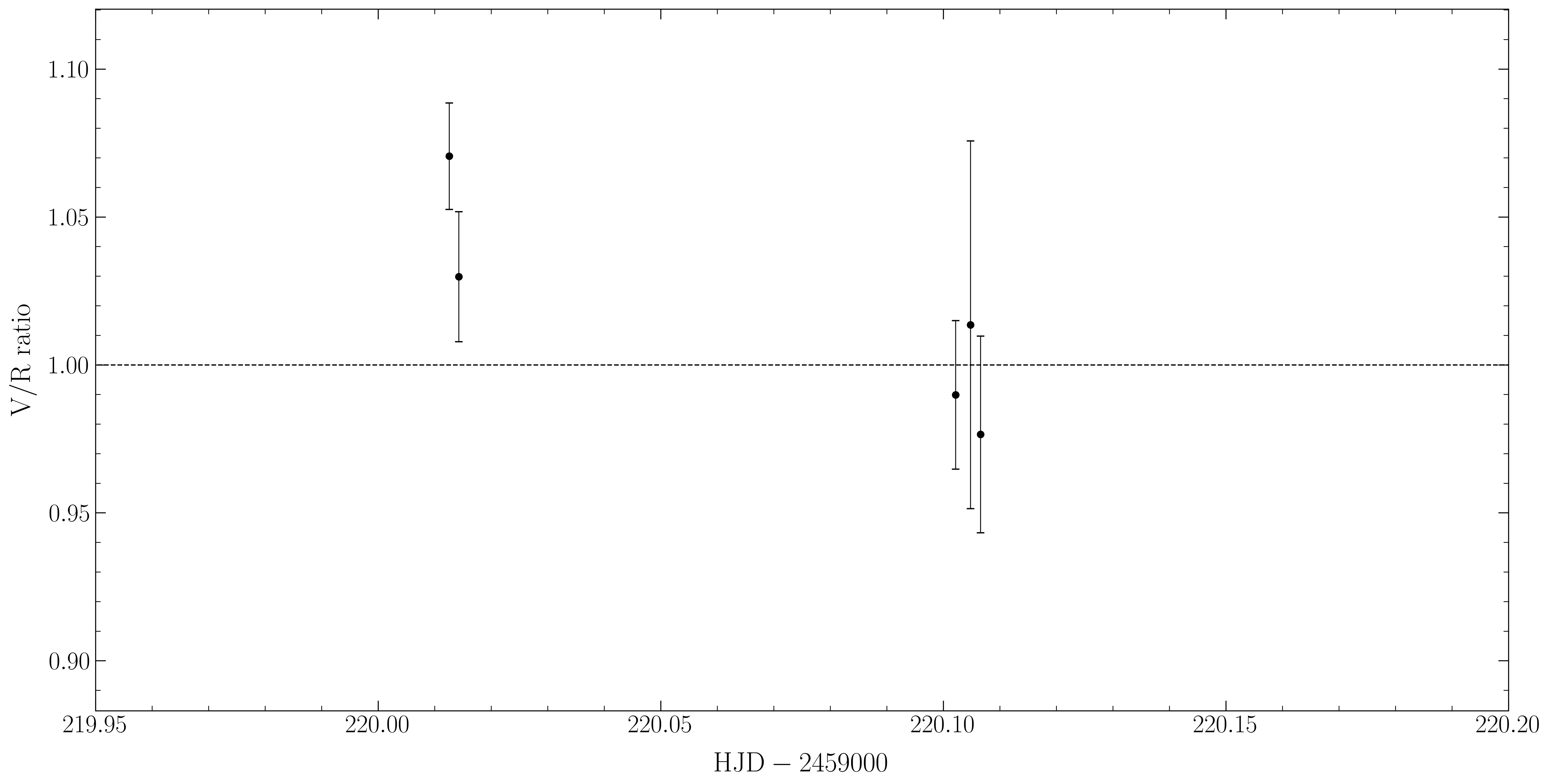}
         \caption{V/R variation in HJD 2459220.}
         \label{fig:220}
     \end{subfigure}
     \hfill
     \begin{subfigure}[b]{\textwidth}
         \centering
         \includegraphics[width=0.9\textwidth]{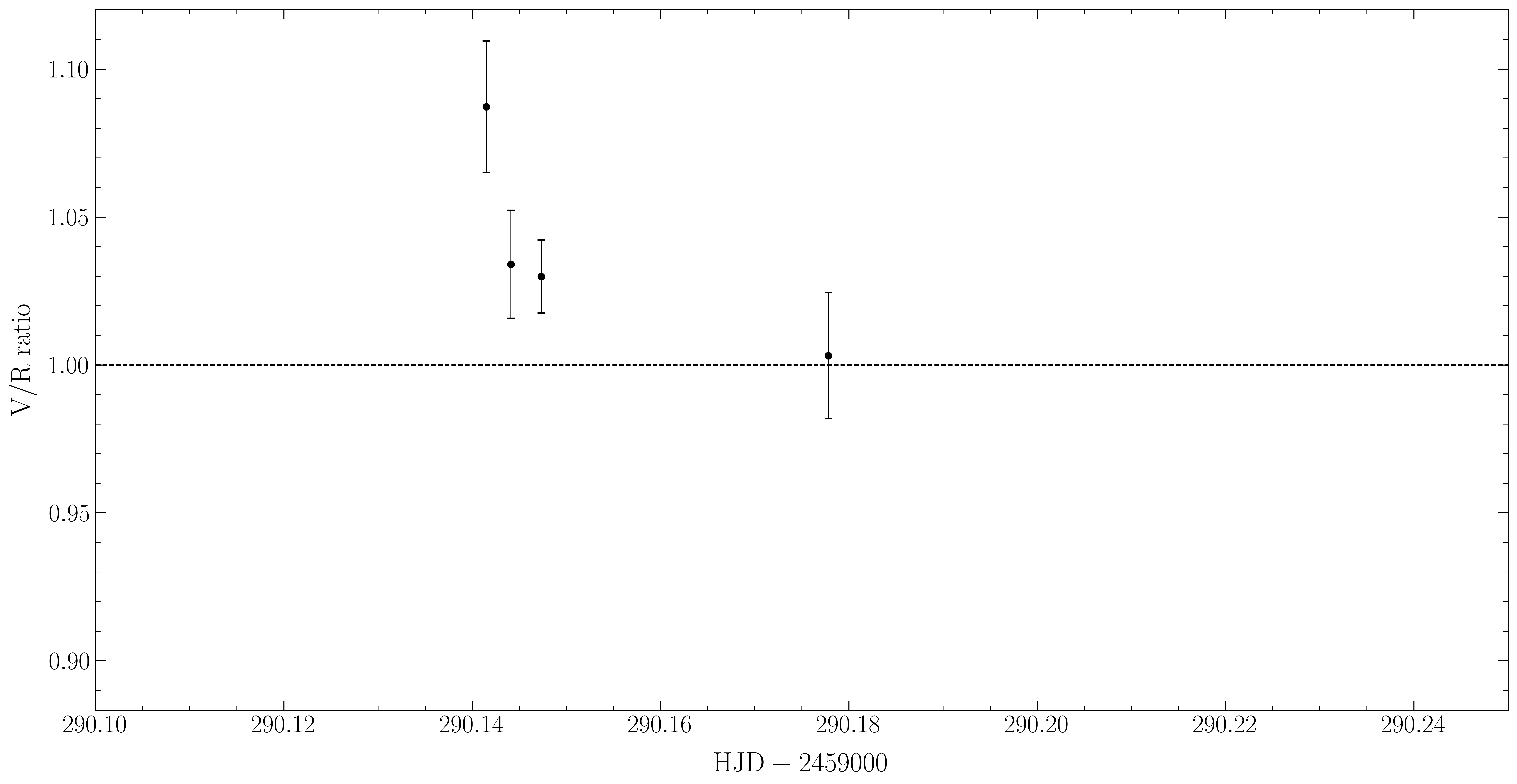}
         \caption{V/R variation in HJD 2459290.}
         \label{fig:290}
     \end{subfigure}
        \caption{Three V/R variations when data allows us to analyze these changes.}
        \label{fig:vr_zoomin}
\end{figure}

In our observation, we can see the V and R variations of H$\beta$ lines, as shown in \textbf{Figure~\ref{Fig:Hbeta_variation}}. It is of great interest to use this emission feature to investigate possible V/R variations that might indicate asymmetries in the circumstellar disc. To this end, the V/R ratio of the H$\beta$ line was measured across the entire observations. The values for V and R correspond to the peak intensities, which were derived through double Gaussian fitting employing the least square method. The uncertainties associated with these measurements are determined through error propagation techniques. The results are depicted in \textbf{Figure~\ref{Fig:vr_variation}} and detailed in \textbf{Table~\ref{Table:VR}}. There is evidence that variability exists in the ratio. The values change from 0.94 to 1.20 over 170 days. Due to the observation period spanning many days, we have observed significant changes over a short period. As a result, we can analyze each observation day in detail, particularly when data allows us to analyze these changes, as shown in \textbf{Figure~\ref{fig:vr_zoomin}}. 

In \textbf{Figure~\ref{fig:172}} corresponding to HJD of 2459172 days, we found that the majority of data points fall below the V$=$R horizontal line, mostly in the V/R$<$1 segment throughout the data period, which spans approximately 189.76 minutes. In \textbf{Figure}~\ref{fig:220}, we can separate data into two datasets. The first dataset decreased rapidly, while the second dataset appeared to fluctuate. The time interval between the first and second data sets is 126.58 minutes, while the overall time interval in \textbf{Figure~\ref{fig:220}} is 135.33 minutes.In \textbf{Figure~\ref{fig:290}}, every point has a V/R ratio greater than 1. The first three points appeared to decrease, while the last is close to V=R. The time interval between the first three points is 8.41 minutes, with the fourth point located 43.90 minutes away from them. The preliminary results suggest that EXO 051910+3737.7 has quasiperiodic less than one day. The trend is inconclusive because the ranges of the error bars are comparable to the variabilities of the data trends. In a previous study, a periodicity of order of 300 seconds for V/R variations was found \cite{Rossi1991}. If so, it is much more challenging to determine the quasiperiodic phase of this object with current exposure times due to the limitation of the telescope. Given the quasiperiodic phase observed in previous studies with a duration of 300 seconds, considering the signal-to-noise ratio (S/N), we are constrained not to reduce the exposure time below 60 seconds and the readout time of approximately 30 seconds. This limitation implies obtaining at most three data points, which is insufficient for analyzing the quasiperiodic phase.

\renewcommand{\arraystretch}{1.5}
\begin{table}
\centering
\caption{The summary of our results which contains the chi-square values ($\chi^2$), analyzed using doubled-Gaussian profile fitting, V/R measurement in which V and R are peak intensities, the half-peak separation ($\Delta\lambda$) observed within the H$\beta$ emission lines and the ratio of radius $r_{\beta}$ of the H$\beta$ emitting envelope over stellar radius ($r_*$).}
\label{Table:VR}
\begin{tabular}{ccccccc}
\hline

\textbf{Date} & \textbf{Frame} & \textbf{HJD} &  \textbf{$\chi^2$} & \textbf{V/R ratio} & \textbf{$\Delta$$\lambda$ (\AA)} & \textbf{$r_{\beta}/r_*$}\\ 
\hline
18--Nov--2020   & 001 & 2459172.31073 & 0.055  & 0.910$\pm$0.016 & 3.109$\pm$1.482 & 2.603$\pm$1.243\\
                & 002 & 2459172.31291 & 0.035  & 0.932$\pm$0.013 & 3.085$\pm$1.490 & 2.623$\pm$1.268\\
                & 003 & 2459172.37204 & 0.029  & 0.943$\pm$0.013 & 3.111$\pm$1.517 & 2.601$\pm$1.269\\
                & 004 & 2459172.44048 & 0.028  & 0.942$\pm$0.013 & 3.090$\pm$1.498 & 2.619$\pm$1.271\\
                & 005 & 2459172.44251 & 0.033  & 0.966$\pm$0.014 & 3.096$\pm$1.468 & 2.614$\pm$1.241\\
\hline
02--Jan--2021   & 001 & 2459217.13162 & 0.096  & 1.022$\pm$0.031 & 3.118$\pm$1.451 & 2.595$\pm$1.209\\
                & 002 & 2459217.13351 & 0.133  & 0.941$\pm$0.032 & 3.240$\pm$1.507 & 2.498$\pm$1.164\\
                & 003 & 2459217.13541 & 0.041  & 1.020$\pm$0.017 & 3.121$\pm$1.448 & 2.593$\pm$1.204\\
\hline
05--Jan--2021   & 001 & 2459220.01260 & 0.040  & 1.071$\pm$0.018 & 3.085$\pm$1.474 & 2.623$\pm$1.255\\
                & 002 & 2459220.01428 & 0.062  & 1.030$\pm$0.022 & 3.082$\pm$1.463 & 2.626$\pm$1.248\\
                & 003 & 2459220.10218 & 0.090  & 0.990$\pm$0.025 & 3.148$\pm$1.417 & 2.571$\pm$1.158\\
                & 004 & 2459220.10482 & 0.312  & 1.014$\pm$0.062 & 3.313$\pm$1.472 & 2.443$\pm$1.087\\
                & 005 & 2459220.10658 & 0.170  & 0.976$\pm$0.033 & 3.127$\pm$1.458 & 2.589$\pm$1.209\\
\hline
16--Mar--2021   & 001 & 2459290.14150 & 0.050  & 1.087$\pm$0.022 & 3.108$\pm$1.428 & 2.604$\pm$1.197\\
                & 002 & 2459290.14412 & 0.050  & 1.034$\pm$0.018 & 3.138$\pm$1.491 & 2.579$\pm$1.226\\
                & 003 & 2459290.14734 & 0.023  & 1.030$\pm$0.012 & 3.124$\pm$1.466 & 2.591$\pm$1.217\\
                & 004 & 2459290.17782 & 0.061  & 1.003$\pm$0.021 & 3.173$\pm$1.486 & 2.550$\pm$1.196\\
\hline
18--Mar--2021   & 001 & 2459292.04618 & 0.030  & 1.052$\pm$0.014 & 3.134$\pm$1.503 & 2.583$\pm$1.240\\
\hline
08--Apr--2021   & 001 & 2459313.02591 & 0.030  & 1.052$\pm$0.014 & 3.109$\pm$1.497 & 2.603$\pm$1.255\\
                & 002 & 2459313.05456 & 0.046  & 1.031$\pm$0.017 & 3.070$\pm$1.500 & 2.636$\pm$1.289\\
\hline
07--May--2021   & 001 & 2459342.03902 & 0.173  & 0.966$\pm$0.030 & 3.261$\pm$1.500 & 2.482$\pm$1.143\\
                & 002 & 2459342.04412 & 0.069  & 1.034$\pm$0.021 & 3.069$\pm$1.463 & 2.637$\pm$1.258\\
\hline
\end{tabular}
\end{table}
\renewcommand{\arraystretch}{1}

\section*{\normalsize Radius $r_{\beta}$ of the H$\beta$ emitting envelope} 
The estimation of the H$\beta$ envelope's radius can be derived from the half-peak separation ($\Delta\lambda$) in km\,s$^{-1}$ observed within the H$\beta$ emission lines \cite{mennickent1991}. This parameter can be calculated using the following equations:

\begin{equation}
\Delta\lambda = v_{\beta}\sin i,
\end{equation}
The above equation is applied to the case of an envelope shape like a disc with an outer rotational velocity $v_{\beta}$ \cite{Hirata1984}. The rotation of the envelope is specified as follows:

\begin{equation}
v(r) = \left(\frac{v_0}{r/r_*}\right)^j,
\end{equation}
where $v_0$ is the critical velocity and is defined as follows:
\begin{equation}
v_0 = \sqrt{\frac{Gm_*}{r_*}},
\end{equation}
where $j$ is a rotation exponent. Under these conditions, the expression for $r_{\beta}$ is given by,

\begin{equation}
\frac{r_{\beta}}{r_*} = \left(\frac{v_{\beta}}{v_0}\right)^{-1/j} = \left(\frac{\Delta\lambda \, v}{v \sin i \, v_0}\right)^{-1/j} .
\end{equation}

We adopted the assumption that $v=$ 350 km\,s$^{-1}$, which is a reasonable approximation for the rotational velocity commonly observed in Be stars displaying double-peak emission lines \cite{mennickent1991}. Furthermore, we employed a $j$ value of approximately 1, consistent with the analysis by Mennickent et al. (1991) \cite{mennickent1991}. The $v_0$ values in our analysis correspond to those typical of normal B-type stars \cite{Doazan1982}. We used $m_*$ and $r_*$ values as 16.7$\pm$0.4 $M_{\odot}$ and 6.0$\pm$0.1 $R_{\odot}$, respectively, based on the object's spectral type, which is B0IV for our target. We also used the rotational velocity ($v \sin i$) of 240$\pm$4 km\,s$^{-1}$ \cite{Halbedel1996}. From our observed H$\beta$ spectra, we obtained wavelength difference at the peak ($\Delta\lambda$) in km\,s$^{-1}$ from two separate peaks. This $\Delta\lambda$ was obtained by fitting the line profile using the methodology mentioned in the Line Profile Fitting section and shown in \textbf{Table~\ref{Table:VR}}. Then, we convert the wavelength differences to velocities using the Doppler effect as given by,

\begin{equation}
\frac{\Delta\nu}{\lambda_0} = \frac{\Delta\lambda}{c},
\end{equation}
where $\lambda_0$ is the rest wavelength of the H$\beta$ line at 4861.333$\pm$0.003 \AA  \cite{nist2022} and the velocity differences. The speed of light in a vacuum is denoted by $c$. These parameters were employed to estimate the $r_{\beta}$ values and their associated uncertainties as presented in \textbf{Table~\ref{Table:VR}}. The average $r_{\beta}$ is 2.585$\pm$0.050$r_*$.

\section*{\normalsize Conclusions}

In this study, we have presented 22 spectral line profiles of observed over seven nights during the 2020-2021 period using the TNO/MRES instrument. Despite the limitations imposed by medium-low resolution, metallic lines were not discernible. However, our observations did reveal the presence of absorption lines for He I $\lambda$6678 and He I $\lambda$4922. Notably, H$\alpha$ and H$\beta$ lines exhibited the characteristic double-peak emission lines associated with Be stars. By applying Gaussian models for fitting the H$\beta$ spectral line, we were able to derive essential parameters and associated uncertainties, including the V/R ratio and $\Delta\nu$, leading our investigation to the physical characteristics of EXO 051910+3737.7. Our examination of V/R variation found rapid changes occurring within a single day. However, estimating the periodicity of these variations was challenging due to the limitations of the instrument and the duration of our observations. 

Furthermore, we can estimate the radius, denoted as $r_{\beta}$, for the H$\beta$ emitting envelope by utilizing the observed $\Delta\nu$ and typical Be star parameters. The average value of $r_{\beta}$ was determined to be 2.585$\pm$0.050$r_*$. In future investigations, we plan to employ high-resolution instruments and larger telescopes to obtain more detailed and comprehensive data and to further enhance our understanding of this object.

\section*{\normalsize Acknowledgements}
The authors would like to thank the referees for their valuable comments and suggestions which helped to improve the manuscript. This work has made use of data based on observations made with MRES at the Thai National Observatory under program ID 20 in 2020 and ID 04 in 2021, which is operated by the National Astronomical Research Institute of Thailand (Public Organization). PN is partially supported by funding from Thai Government’s Development and Promotion of Science and Technology Talents Project (DPST), the National Astronomical Research Institute of Thailand (NARIT) and the University of Southampton.


\end{document}